# A Scalable All-to-All Reconfigurable Ising Solver Using Pulsed Time-Division Multiplexing

Henry Love[1,†], Zhehao Yu[1,†], Mohammad Alimadadi[1], Yan Jin[2,3], Nitesh Chauhan[2,3], Brian Edwards[1], Pratik Chaudhari[1], Scott B. Papp[3], and Firooz Aflatouni[1,*]

[1] Department of Electrical and Systems Engineering, University of Pennsylvania, Philadelphia, PA, USA
[2] Department of Physics, University of Colorado, Boulder, CO, USA
[3] Time and Frequency Division, National Institute of Standards and Technology, Boulder, CO, USA
[†] Authors with equal contributions
* firooz@seas.upenn.edu

## Abstract

**Physics-based computing platforms, such as those based on the Ising model, are an important pillar of future hardware systems built for the artificial intelligence (AI) era. Such platforms[1-15] show promise for solving nondeterministic polynomial (NP) time problems that are difficult for traditional processing units to solve efficiently as problem size grows[16,17]. Here, we present a scalable optoelectronic Ising machine architecture, demonstrated with 64 all-to-all connected spins using pulsed time-division multiplexing. The 65 nm CMOS Ising chip integrates the coupling and nonlinear mechanisms in an active area of 3.1 mm$^2$, eliminating the need for benchtop equipment within the loop. The feedback loop of the Ising machine is closed using a compact high-bandwidth, low-loss optical fiber, seamlessly combining optical scalability with the ultradense reconfigurability of integrated electronics. The chip operates at 1 GHz with 4-bit coupling weights and is benchmarked with NP-complete Boolean satisfiability problems consisting of three literals (3-SAT) and clause-to-variable ratios of 32/32, 40/24, and 48/16. Nanosecond annealing times represent at least a three order-of-magnitude improvement over previously reported all-to-all connected works. Time and energy to solutions for 100% 3-SAT clause accuracy are as low as 7.4 $\mu$s and 2.9 $\mu$J, respectively, achieving more than an order-of-magnitude decrease in time and energy to solution compared to the state of the art. All-to-all connection is demonstrated using MaxCut problems with 100% graph densities. The chip's ability to effectively solve 2-, 3-SAT, and MaxCut problems highlights its reconfigurability and versatility. Furthermore, combining mature CMOS integration with scalable photonic links allows for significant reduction in computation time and energy, addressing the pressing requirements of AI and future hyperscale datacenters[18-22].**

## Introduction

Combinatorial optimization problems (COPs) find applications across artificial intelligence (AI), route planning, drug discovery, network design, and robotics[23]. These problems are characterized by grouping discrete variables in an exponentially growing solution space, placing many of them in the class of nondeterministic polynomial (NP) time problems[16,17]. Conventional computing platforms that use central processing units (CPUs), graphics processing units (GPUs), and field-programmable gate arrays (FPGAs) are highly versatile due to their precise, temporally stable, noise-immune, programmable digital memory architectures[24] but have struggled when applied to large COPs. While these platforms can solve moderate-scale COP instances using sophisticated heuristic algorithms[25-27], their efficiency degrades rapidly as problem size increases, resulting in prohibitive time to solution and energy consumption[28]. This inefficiency stems from a fundamental mismatch between the inherently parallel, strongly coupled structure of COPs and the sequential control flow and memory-centric operation of von Neumann architectures.

To overcome this mismatch, a growing class of physics-inspired computing approaches seek to solve COPs by embedding them into the dynamics of physical systems that naturally evolve toward low-energy states. In such inherently parallel energy-based formulations, optimization is performed through collective relaxation rather than explicit algorithmic iteration. A prominent example of this paradigm is the Ising model[29]. In this model, binary decision variables are represented by interacting spins, and the solution to the optimization problem corresponds to the ground state of an associated Hamiltonian[30]. Ising machines aim to find a configuration of bistable spins $\sigma_i \in \{+1, -1\}$ that minimize the Ising energy, or equivalently, the Ising Hamiltonian, written as

$$H = -\boldsymbol{\sigma}^T \mathbf{J} \boldsymbol{\sigma} - \mathbf{h}^T \boldsymbol{\sigma} = -\sum_{i,j} \mathrm{J}_{ij} \sigma_i \sigma_j - \sum_i \mathrm{h}_i \sigma_i \,, \tag{1}$$

where $\mathbf{J} \in \mathbb{R}^{N\times N}$, and $\mathbf{h} \in \mathbb{R}^{N}$. This natural process offers the prospect of faster convergence and lower energy consumption compared to traditional computing implementations that mimic/simulate the annealing process artificially[27,31]. Key figures of merit for any hardware Ising solver include the level of spin connectivity and coupling, coupling resolution, time to solution, and energy to solution. If each spin couples to every other spin, the system is considered fully connected, or equivalently, has all-to-all coupling. Such fully connected systems pose significant scaling challenges as the number of spins, $N$, increases.

Over the past decade, several physical platforms have been explored to realize analog Ising solvers to address the demands for scalable, efficient computing architectures for the AI era. Complementary metal-oxide-semiconductor (CMOS) Ising machines allow for mass manufacturability, reliability, and a high level of integration offered by commercial foundries[1-5]. Recent advances include the utilization of CMOS ring-oscillators where spins are encoded in oscillator phases and couplings are realized through electrical connections[3-5]. These platforms benefit from monolithic integration and compact footprints. However, because coupling is implemented through largely hardwired copper interconnect, realizing dense and reconfigurable all-to-all coupling becomes increasingly complex as network size grows and is often not compatible with extreme scalability[32]. As a result, many implementations adopt structured or sparse topologies instead of all-to-all connectivity. For example, in an impressive work, a 560-spin probabilistic compute fabric has been demonstrated[3], where despite supporting a relatively large number of spins, the connectivity is limited to a hexagonally coupled array and 1-bit coupling resolution. At the same time, architectures that support all-to-all coupling have been reported at more modest spin counts. For instance, a 48-node all-to-all coupled-ring-oscillator-based solver has been shown[4], where despite excellent performance, achieving higher weight resolution requires merging available spin cells, which could directly reduce the available spin count under a fixed area and power budget.

To scale to higher spin counts and connectivity simultaneously, optical Ising machines have been proposed that leverage the scalable nature of high-bandwidth and low-loss optical interconnects. Such optical Ising machines generally encode the spins in the phase[6-9] or amplitude[10-12] of the optical carrier. Both categories typically use time-division multiplexing in a fiber cavity to store and update spins as a pulse train circulates. Encoding spins in the phase domain requires precise phase alignment between spins, whereas amplitude-based machines typically utilize the intensity of pulses to encode each spin and are more immune to phase variations in the optical carrier. Phase-domain optical Ising machines can implement 100,000 spins with all-to-all and high-resolution couplings[7]. However, this scalability is achieved with substantial system-level overhead: such machines require long, phase-sensitive optical cavities (e.g., kilometer-scale fiber delay), homodyne readout, low-noise and narrow-linewidth pump lasers, and active stabilization to maintain coherence. Thus, these systems are often implemented using bulky components and suffer from a large wall-plug energy consumption. Machines that encode the spin in the amplitude domain relax the requirements of coherent optics; however, spin coupling and feedback paths are still typically performed by large, benchtop FPGA/host-electronics whose resource and power demands scale unfavorably with dense connectivity[10,11]. Consequently, size, energy, and environmental sensitivity dominated by the benchtop equipment within the loop remain major barriers to compact, deployable optical-based Ising machine hardware with a high energy efficiency. Recent work monolithically integrates coupling and nonlinear blocks to reduce footprint and improve per-spin efficiency, but despite impressive results, current demonstrations remain at the few-spin (e.g., 4-spin) level[12], leaving open the challenge of scaling to large, reconfigurable, all-to-all connectivity on chip.

Other emerging Ising machine implementations leverage a diverse set of material and device platforms including $VO_2$-based phase-transition oscillators[13], superparamagnetic tunnel junctions (SMTJs)[14], and single-photon avalanche diodes (SPADs)[15]. These approaches exploit abrupt and highly nonlinear switching dynamics inherent to the underlying devices to emulate bistable spin states, while intrinsic device noise or stochastic event generation enables probabilistic state updates. In some demonstrations, dense or even all-to-all coupling is realized; however, most reported works achieve coupling through global interactions mediated by external control circuits such as printed circuit board (PCB)-level digital-to-analog converters (DACs), microcontroller units (MCUs), and FPGAs. Therefore, such platforms also face substantial challenges in large-scale integration and system-level wall-plug energy efficiency.

Here, we demonstrate a miniature all-to-all coupled, scalable, electro-optic Ising machine where a CMOS chip realizes low-power, ultradense reconfigurable spin couplings and a low-cost, ultralow-loss, compact optical delay line facilitates system scalability via pulsed time-division multiplexing. The use of amplitude-only optical processing eliminates the need for active fiber phase stabilization yet enables scalability to extreme spin counts with a small overall system volume and without using any benchtop devices. Operating at 1 GHz, the proposed Ising chip is benchmarked with NP-complete Boolean satisfiability problems consisting of three literals (3-SAT) and clause-to-variable ratios of 32/32, 40/24, and 48/16. Time and energy to solutions for 100% 3-SAT clause accuracy are as low as 7.4 $\mu$s and 2.9 $\mu$J, respectively. 2-SAT and MaxCut problem solving is also demonstrated to highlight the system's versatility. Without relying on high quality factor or high-finesse resonators, cavities, or oscillators in the loop, the machine is unconstrained by the large time constants that accompany these structures, achieving ultrafast operation with very high energy efficiency. As a result, the implemented electro-optic Ising solver demonstrates an annealing time which is at least three orders-of-magnitude lower than previously reported all-to-all connected works[1,5,7,9,11,14] and achieves more than an order-of-magnitude improvement in time and energy to solutions compared to other state-of-the-art all-to-all Ising solvers. The proposed architecture paves the way for the implementation of compact, ultralow-energy and ultrafast optimizers for future AI and compute systems.

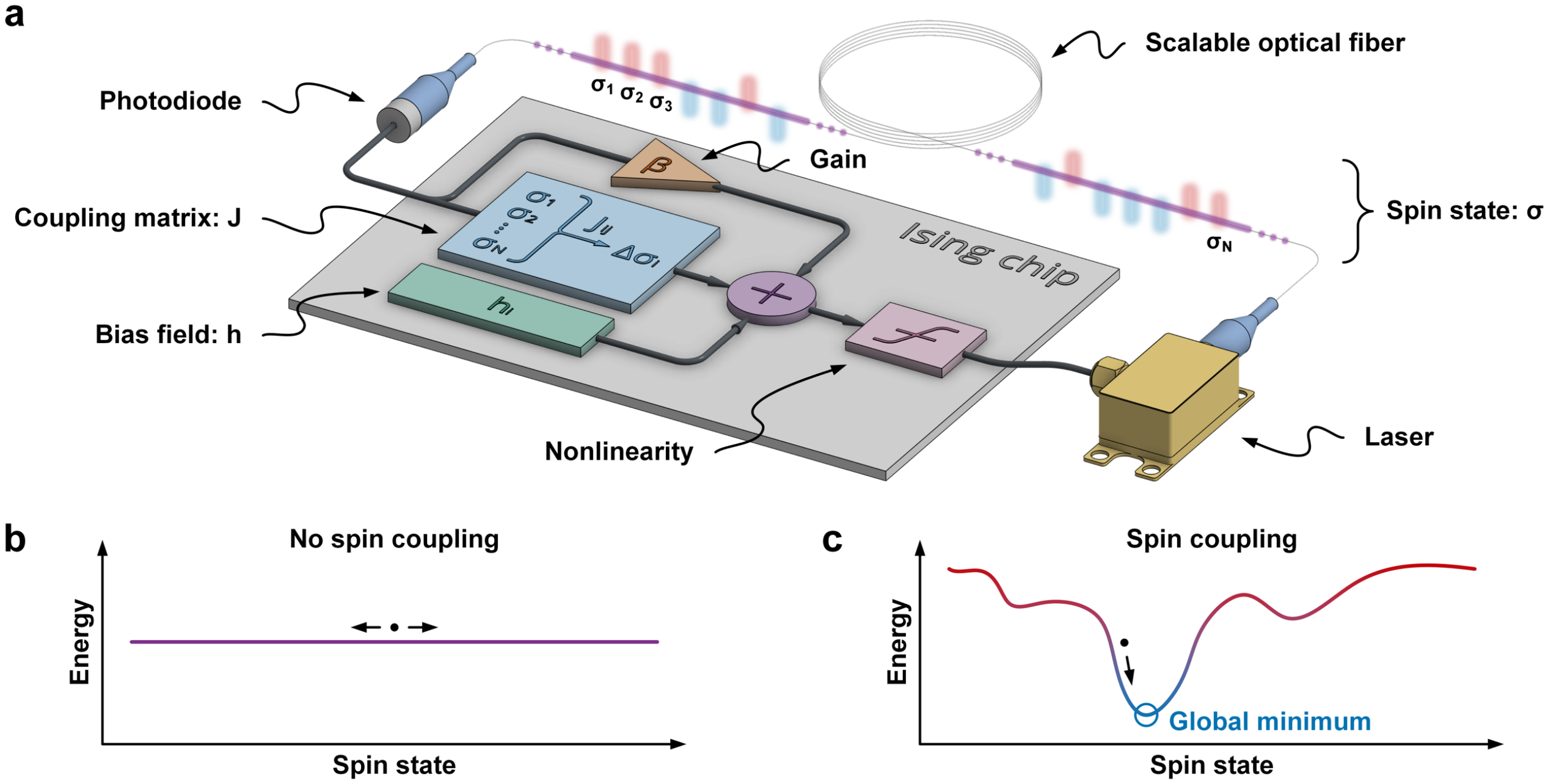

**Fig. 1 | Conceptual diagram of the implemented compact photonic-assisted Ising machine. a** The proposed system comprising of a monolithic Ising chip with feedforward gain $\beta$ (orange), all-to-all spin coupling $\mathbf{J}$ (blue), bias field $\mathbf{h}$ (green), and nonlinearity (pink). A laser and photodiode are used to store pulsed optical spins $\boldsymbol{\sigma}$ in a scalable optical fiber. **b** Energy landscape without spin coupling or bias field. **c** Energy landscape with spin coupling and bias field.

## Results

### A scalable all-to-all reconfigurable Ising machine

The general concept of the proposed photonic-assisted Ising machine is illustrated in Fig. 1a where the reconfigurable monolithic CMOS integrated circuit is responsible for guiding the system to a low-energy state based on the Ising model[29]. The chip includes a feedforward path with a gain of $\beta$ (orange), all-to-all coupling matrix $\mathbf{J}$ (blue), bias field $\mathbf{h}$ (green), and nonlinear output driver (pink). To facilitate scalability, a laser is pulsed using the chip output where the spins $\{\sigma_i \mid i = 1, \ldots, N\}$ for $N$ = 64 are stored in a low-loss, high-bandwidth, optical delay line.

Without coupling between spins (i.e. $\mathbf{J} = \mathbf{0}$) or bias field (i.e. $\mathbf{h} = \mathbf{0}$), a closed-loop system with sufficiently high loop gain will spontaneously evolve based on initial random noise of the system. Eventually, the loop gain will saturate and bifurcate each spin, $\sigma_i$, to $\{+1, -1\}$. Under this condition, there are $2^{64}$ spin vectors, $\boldsymbol{\sigma}$, that are equally favorable due to the flat energy landscape of the bifurcated system, as depicted in Fig. 1b. By introducing a coupling matrix, $\mathbf{J}$, and bias field, $\mathbf{h}$, the energy landscape is contoured, as shown in Fig. 1c, and the system is enticed to relax into a low-energy state. In this case, $\mathbf{J}$ defines spin-to-spin interaction, and the biasing vector, $\mathbf{h}$, adds a fixed term (i.e. bias), $\mathrm{h}_i$, to each spin, $\sigma_i$. After allowing the system to evolve based on the energy landscape, the final spin state, $\boldsymbol{\sigma}$, can be read as a potential solution to the problem mapped onto the machine.

The architecture of the Ising chip is shown in Fig. 2a. A 1:64 interleaver (yellow) is used to store each spin value locally via on-chip sampling capacitors. Each sampling capacitor is connected to a multiplier (blue) which applies a signed weight to the spin voltage stored on each capacitor. This multiplier array is programmed with the weights derived from the coupling matrix, $\mathbf{J}$, where each multiplier in the array has a local memory depth of 64 x 4 bits, corresponding to a 4-bit coupling weight, $\mathrm{J}_{ij}$, spanning from −7 to +7 for every spin pair, $(\sigma_i, \sigma_j)$. To match the delays of the $\mathbf{J}\boldsymbol{\sigma}$ and $\mathbf{h}$ vector paths (Fig. 2a), a multiplier, identical to the ones used in the $\mathbf{J}\boldsymbol{\sigma}$ path but with a constant input, is used to implement $\mathbf{h}$ (green). Like the $\mathbf{J}$ matrix, the multiplier for the $\mathbf{h}$ vector has a local memory depth of 64 x 4 bits, corresponding to a signed 4-bit $\mathrm{h}_i$ for every spin, $\sigma_i$. A feedforward path is also included (orange) which amplifies the current sampled spin value by $\beta$. A current-mode adder (purple) sums the output of the feedforward, $\mathbf{J}\boldsymbol{\sigma}$, and $\mathbf{h}$ vector paths. In this way, the amplified spin value, $\beta\sigma_i$, is perturbed by the $\mathbf{J}$ coupling matrix and $\mathbf{h}$ vector such that the Ising energy is minimized via gradient descent. In this case, the gradient descent learning rate is defined as the gain of the derivative of the Hamiltonian normalized to the feedforward path gain, $\beta$. For more details regarding the optimization algorithm, see the Methods section and Eqns. (5) – (7). More details on the analog signal path and memory architecture of the chip are included in the Methods section and Extended Data Figs. 1 and 2. The last stage of the chip is a saturable output driver (pink) which is pulsed at 1 GHz with a 25% duty cycle to lower the power consumption of the system. The intrinsic nonlinearity of the output driver gives rise to spin bifurcation, allowing the solver dynamics to be implemented without external nonlinear materials. More details on the system nonlinearity are included in the Methods section.

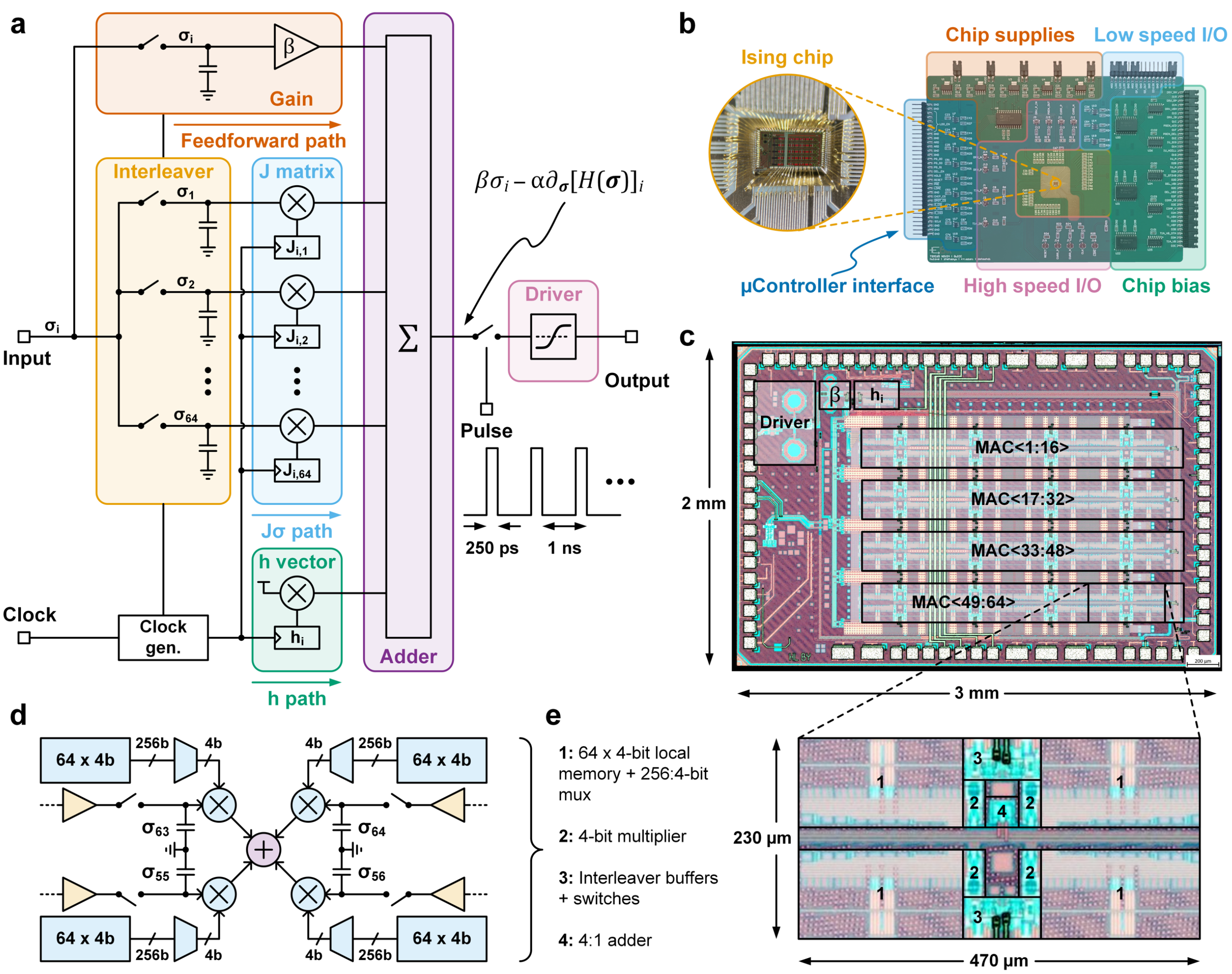

**Fig. 2 | Ising chip. a** Block diagram of the chip architecture. **b** Printed circuit board used for measurement with zoomed-in photograph of the wirebonded custom integrated chip. **c** Micrograph of the Ising chip. **d** Circuit diagram of a unit 4:1 MAC cell which multiplies 4 spins by their corresponding 4-bit coupling values and adds them together using a 4:1 adder. **e** Corresponding layout of 4:1 MAC cell with labels shown on the left.

The PCB for the chip is shown in Fig. 2b, right, with a zoomed-in photograph of the wirebonded integrated circuit shown on the left. The integrated circuit is fabricated on TSMC's 65 nm MS RF GP CMOS process, and the associated chip micrograph is shown in Fig. 2c. The total active area of the chip is approximately 3.1 mm$^2$ with roughly 2.9 mm$^2$ of the area used for the 1:64 interleaver and multiplication and accumulate (MAC) operations required for the all-to-all spin coupling. The matrix multiplication is processed in groups of four, where the circuit architecture of a unit 4:1 MAC cell is shown in Fig. 2d. The corresponding layout of the 4:1 MAC cell is shown in Fig. 2e, where 16 of these unit cells are used to perform the entire $\mathbf{J}\boldsymbol{\sigma}$ matrix multiplication.

## Measurement setup

The measurement setup of the proposed Ising machine is shown in Fig. 3a. Due to the inherent benefits of optical signaling such as high bandwidth, high signal confinement, and low loss[24,33,34], an optical delay line (red) is used to close the loop. The output of the Ising chip drives a 50 Ω, 6.5 GHz differential readout buffer where one output modulates a 1550 nm distributed feedback (DFB) laser with integrated electro-absorption (EA) modulator, and the other output is used for readout and monitoring. To time-synchronize the output pulse of the chip with the sampling point of the input interleaver, the output of the chip is delayed by 64 ns (~12.8 m) of single-mode optical fiber with loss of 0.19 dB/km at $\lambda$ = 1550 nm. A variable optical delay line with continuous tunable range of 600 ps is used as fine adjustment for the optical delay. A co-packaged 25 GHz photodiode (PD) and transimpedance amplifier (TIA) is used to convert the optical pulse train back to the electrical domain for processing by the Ising chip. The system is clocked at 1 GHz such that one round-trip iteration of all-to-all spin coupling takes 64 ns. An on-chip frequency divider is used to synchronize the output waveform with the readout of the Ising machine.

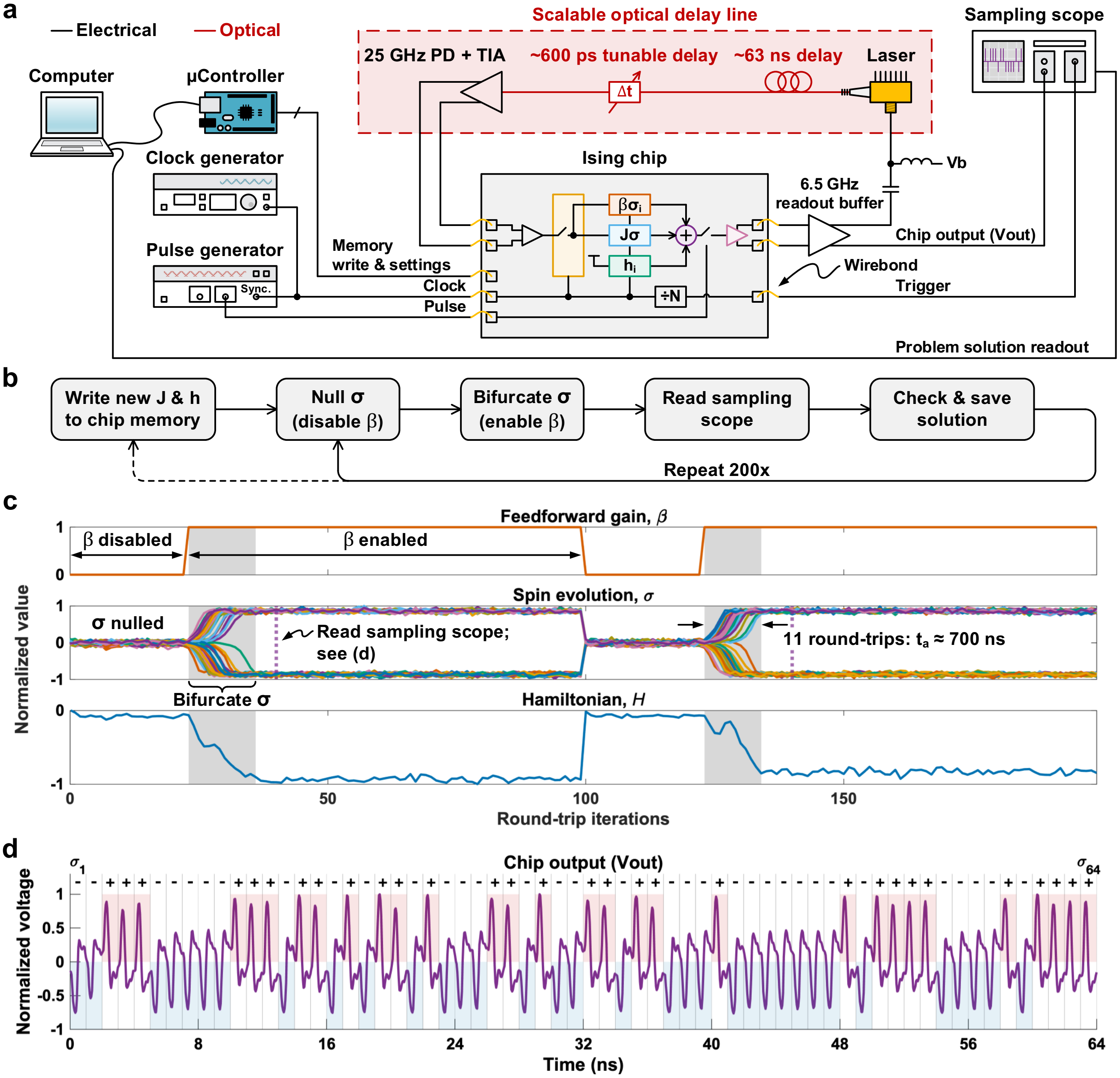

**Fig. 3 | Measurement setup for solving Ising problems. a** The measurement setup and **b** associated state machine diagram used to characterize the Ising system. **c** Simulated evolution of $\boldsymbol{\sigma}$ and $H$ as $\beta$ is disabled/enabled. **d** An example measurement of the final spin state read from the sampling oscilloscope, where a positive pulse represents a positive spin (red) and a negative pulse represents a negative spin (blue). The sampling oscilloscope is triggered by an on-chip clock divider and appropriately delayed ensuring the first spin corresponds to $\sigma_1$.

The diagram of the state machine used for solving each Ising problem is shown in Fig. 3b. First, the on-chip memory is programmed with the $\mathbf{J}$ matrix and $\mathbf{h}$ vector corresponding to the problem to be solved. To begin, the on-chip feedforward gain, $\beta$, is disabled, such that $\boldsymbol{\sigma} = \mathbf{0}$, i.e. there is insufficient loop gain to sustain a non-zero spin vector. With the $\mathbf{J}$ matrix and $\mathbf{h}$ vector enabled, the feedforward gain is increased (Fig. 3c, top) and the spins evolve based on gradient descent (shaded region of Fig. 3c). The simulated evolution of the spins for a 64-spin problem is shown in Fig. 3c (middle), where, on average, it takes approximately 11 round-trip iterations ($t_a \approx 700$ ns) for every spin to bifurcate, i.e. anneal (see the Methods section and Extended Data Fig. 7 for more details). The corresponding normalized Hamiltonian, shown in Fig. 3c (bottom), is calculated using the analog values of each spin. Once the system has bifurcated into its final spin state, the time-domain output of the chip is captured by a sampling oscilloscope. An example measured waveform from the chip is shown in Fig. 3d where a peak search algorithm is applied to determine the $\{+1, -1\}$ state of each spin, as highlighted in red and blue, respectively. Once the solution waveform has been captured and saved, $\boldsymbol{\sigma}$ is nulled again (i.e. $\beta$ is disabled), and the process is repeated 200 times per problem to obtain a distribution of solution vectors.

The total dynamic power consumption of the Ising chip while solving a problem is approximately 400 mW. This is attributed to the digital, analog, and driver circuitry with values of 130 mW, 252 mW, and 18 mW, respectively. To reduce power, a passive, electronic delay line could be used in lieu of the optical delay line. However, the power consumptions of the laser and TIA are amortized by higher spin counts as the system scales, since these values remain fixed regardless of the number of spins. The laser in Fig. 3a consumes 70mW. A detailed list of the equipment used is included in Extended Data Fig. 9.

### Solving $k$-SAT and MaxCut problems

The proposed Ising machine was benchmarked using randomly generated Boolean satisfiability (SAT) and MaxCut problems. To demonstrate the reconfigurability of the machine, 300 unique 2-SAT (nondeterministic logarithmic (NL)-complete), 300 unique 3-SAT (NP-complete), and 300 unique MaxCut (NP-complete) problems were solved.

### $k$-SAT

A Boolean satisfiability problem is defined by a set of clauses where each clause is a logical expression of $k$ literals. Each satisfiability problem consists of $n$ variables $x_i \in \{\text{True}, \text{False}\}$ and $m$ clauses $c_i \in \{\text{True}, \text{False}\}$, where a clause consists of logical disjunction of $k$ variables or their negation. For example, a 4-clause, 5-variable, 3-SAT problem, $S$, may be defined by the conjunction of a set of clauses as

$$S = \underbrace{(\neg x_1 \vee x_3 \vee x_5)}_{c_1} \wedge \underbrace{(x_2 \vee x_3 \vee x_4)}_{c_2} \wedge \underbrace{(x_1 \vee \neg x_3 \vee \neg x_4)}_{c_3} \wedge \underbrace{(\neg x_3 \vee x_4 \vee \neg x_5)}_{c_4}, \quad (2)$$

for a given set of variables $x_1, x_2, x_3, x_4, x_5$. If $S$ evaluates to $\text{True}$, then the problem satisfied. Clause error(s) occur if any of the clauses are not satisfied, i.e. $c_i$ evaluates to $\text{False}$. In this case, clause accuracy may be defined as the ratio of satisfied clauses to the total number of clauses.

The generated $k$-SAT problems were translated into $\mathbf{J}$ and $\mathbf{h}$ using the quadratic unconstrained binary optimization (QUBO) formulation[35]. For 2-SAT problems, each variable directly maps to one of the spins of the system where $x_i \in \{\text{True}, \text{False}\}$ was translated to $\sigma_i \in \{+1, -1\}$. Therefore, a single chip of the proposed hardware can solve $N = n = 64$ variable 2-SAT problems with arbitrary number of clauses, $m$, with the solution space becoming increasingly

**Table 1: Summary of $k$-SAT problems solved on the proposed Ising machine**

| Problem type | Number of clauses ($m$) | Number of variables ($n$) | Number of problems | Number of trials per problem | Total trials | Median clause accuracy (1) |
|---|---|---|---|---|---|---|
| 2-SAT | 64 | 64 | 100 | 200 | 20,000 | 96.4% |
| | 96 | 64 | 100 | 200 | 20,000 | 92.5% |
| | 128 | 64 | 100 | 200 | 20,000 | 90.7% |
| 3-SAT | 32 | 32 | 100 | 200 | 20,000 | 96.9% |
| | 40 | 24 | 100 | 200 | 20,000 | 96.6% |
| | 48 | 16 | 100 | 200 | 20,000 | 91.7% |

(1) Median per-problem mean accuracy

Table 2: Summary of MaxCut problems solved on the proposed Ising machine

| Problem type | J density | Number of variables | Number of problems | Number of trials per problem | Total trials | Median cut value accuracy (1) |
|---|---|---|---|---|---|---|
| MaxCut | 20% | 64 | 100 | 200 | 20,000 | 99.9% |
| | 60% | 64 | 100 | 200 | 20,000 | 96.3% |
| | 100% | 64 | 100 | 200 | 20,000 | 90.7% |

(1) The ratio of the median per-problem mean cut value achieved by the chip and by a digital computer (Apple M4 Macbook Pro) running a standard gradient descent algorithm.

sparse for larger $m$. As summarized in Table 1, of the 300 2-SAT problems, clause-to-variable ratios of $m/n$ = 64/64, 96/64, and 128/64 were solved.

For a 3-SAT problem with $m$ clauses and $n$ variables, a total of $N = m + n$ spins is required. This is due to the cubic nature of the cost function, requiring an ancilla variable per clause. Thus, a single chip of the proposed hardware can solve 3-SAT problems with $N = m + n = 64$ where the first $n$ values of $\boldsymbol{\sigma}$ correspond to the $n$ variables, $x_i$, and $\sigma_i \in \{+1, -1\}$ is interpreted as $x_i \in \{\text{True}, \text{False}\}$. As listed in Table 1, of the 300 3-SAT problems, clause-to-variable ratios of $m/n$ = 32/32, 40/24, and 48/16 were solved.

**MaxCut**

A MaxCut problem is defined as a graph of connected nodes, where the $\mathbf{J}$ matrix defines the weight or cost of each connection. The goal is to partition the nodes into two sets, $\sigma_i \in \{+1, -1\}$, such that when the sets are divided (i.e. cut), the sum of the weights intersected is maximized. MaxCut problems are used to demonstrate the capability of the machine in solving problems represented by not only partial, but also true all-to-all coupling between the spins. The density of the $\mathbf{J}$ matrix corresponds to the level of the coupling between the spins (e.g. a density of 100% represents a true all-to-all coupling). Of the 300 MaxCut problems, $\mathbf{J}$ densities of 20%, 60%, and 100% were solved, where $\mathrm{J}_{ij}$ was randomly chosen between $-7 \leq \mathrm{J}_{ij} \leq 7$ to obtain the desired density. In this case, the $\mathbf{h}$ vector is disabled and the $\mathbf{J}$ matrix is solely responsible for guiding the chip to the solution. A summary of the MaxCut problems run on the machine are shown in Table 2, where median cut value accuracies of the chip are compared to an Apple M4 MacBook Pro[36] solving the same problems (see the following section for further details). In this case, median cut accuracy is the median cut value produced by the chip normalized to the median cut value generated by running gradient descent on the Apple M4 MacBook Pro.

**System performance**

The measured system performance is shown in Fig. 4. For 2-SAT (green) and 3-SAT (blue) problems, the measured clause accuracy is displayed for the different clause-to-variable ratios. Figs. 4a – f show histograms of the results from all 200 trials performed for each SAT problem where the bin width of each histogram corresponds to one clause error; in other words, each histogram has a bin width of $100\%/m$. The mean clause accuracy is then computed for each problem and plotted as a box chart, shown as insets. As a point of reference, given that 2- and 3-SAT clauses are logical disjunctions of 2 and 3 literals, there is a 1/4 and 1/8 probability for a clause to be False for a random trial solution, respectively. The median per-problem mean clause accuracies for 2-SAT are 96.4%, 92.5%, and 90.7% for $m/n$ = 64/64, 96/64, and 128/64, respectively. Additionally, the median per-problem mean clause accuracies for 3-SAT are 96.9%, 96.6%, and 91.7% for $m/n$ = 32/32, 40/24, and 48/16, respectively. Furthermore, out of the 20,000 trials for each $m/n$ = 32/32, 40/24, and 48/16, the machine reached an optimal solution 7087, 4159, and 401 times, respectively, where an optimal solution corresponds to a spin vector, $\boldsymbol{\sigma}$, providing 100% clause accuracy.

The measured system performance for MaxCut problems is shown in Figs. 4g – i where histograms (top) of all 200 trials performed for each MaxCut problem are displayed. The mean cut value is then computed for each problem and plotted as a box chart, as displayed in the bottom of each figure. For the 20%, 60%, and 100% $\mathbf{J}$ densities, the median per-problem mean cut values computed by the chip were 112.4, 189.2, and 273.5, respectively. Given more non-zero $\mathrm{J}_{ij}$ coupling weights for higher $\mathbf{J}$ densities, it is expected that the cut value increases with density.

To benchmark the MaxCut results, the chip solutions were compared to cut values generated by an Apple M4 MacBook Pro running gradient descent on the same MaxCut problems solved on the implemented Ising hardware. For fair comparison, the computer gradient descent hyperparameters (i.e. learning rate and iteration count) were set to match those of the chip. The learning rate of the chip (0.155) was determined by simulating the relative strengths of the feedforward and $\mathbf{J}\boldsymbol{\sigma}$ paths using the same settings applied in measurement. For the computer, total gradient descent iterations were set to 13, 8, and 7 for 20%, 60%, and 100% $\mathbf{J}$ densities, respectively, where these numbers of round-trip iterations fall within the range of the histograms shown in Extended Data Figs. 7a – c. Note that for higher $\mathbf{J}$ densities, the $(\mathbf{J}\boldsymbol{\sigma})_i$ coupling term is larger, likely causing the spins to bifurcate sooner with fewer number of round-trip iterations. Like the system measurements, all 300 unique MaxCut problems were each run 200 times on the computer. For each trial on the computer, random noise was added to the initial condition to allow for a distribution of results. The comparison of the chip and the computer results are shown in Figs. 4g – i (bottom) and summarized in Table 2. In this case, the chip obtains similar cut values as the computer for 20% $\mathbf{J}$ density. For 60% and 100% $\mathbf{J}$ densities, the chip obtains median per-problem mean cut values 96.3% and 90.7% of the computer results, respectively. To compute a single cut value on the Apple M4 MacBook Pro, it takes approximately 25 $\mu$s per trial at a power consumption of approximately 3 W[36]. Thus, the annealing time and energy of the computer are approximately 25 $\mu$s and 75 $\mu$J, respectively. Comparing these values to those of the chip (i.e. annealing time and energy of 700 ns and 280 nJ, respectively), the chip offers two orders-of-magnitude improvement in both annealing time and energy compared to the digital computer.

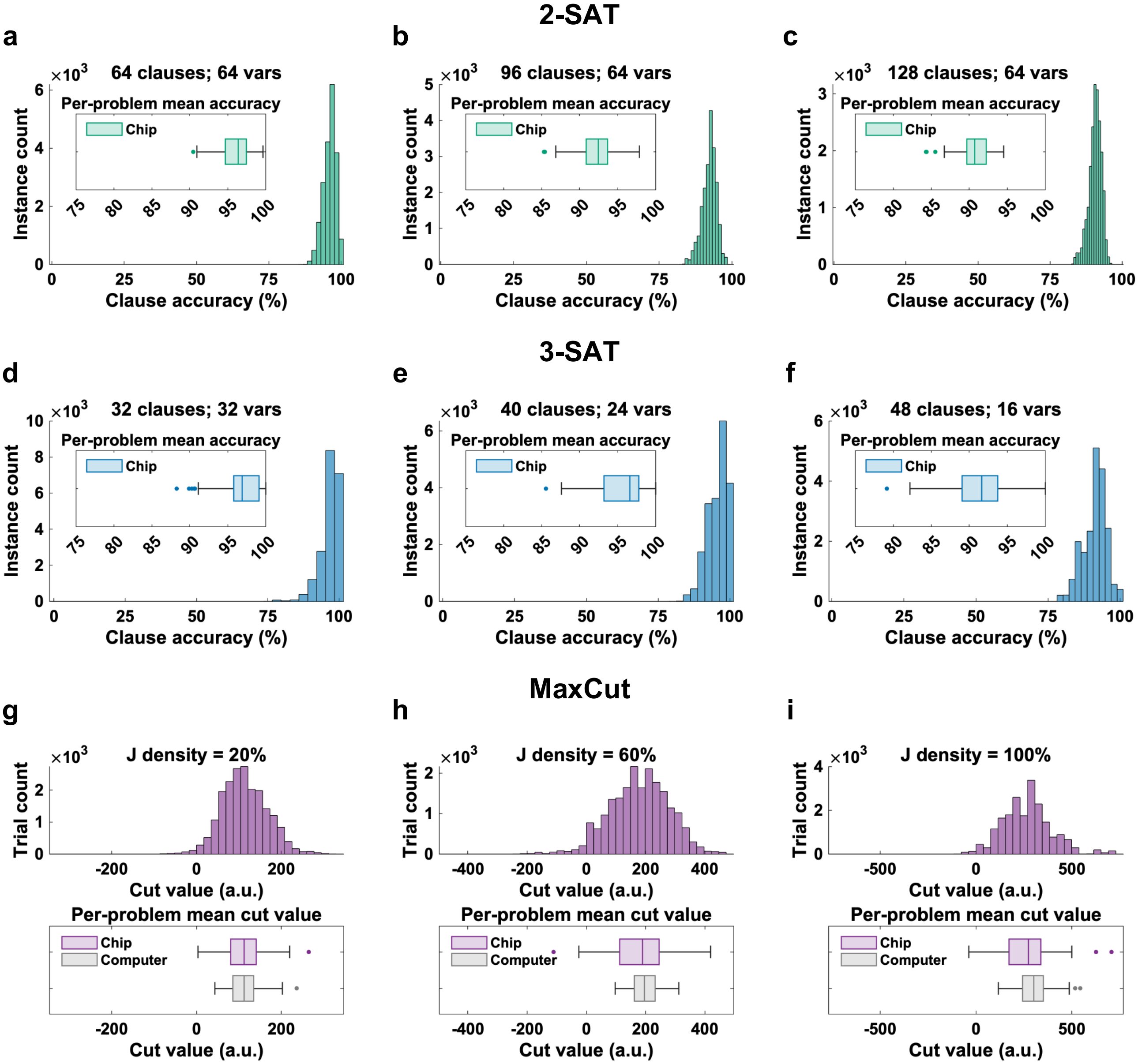

**Fig. 4 | Demonstration of solving $k$-SAT and MaxCut problems.** Clause accuracy of 2-SAT problems (green) with clause-to-variable ratios, $m/n$, of **a** 64/64, **b** 96/64 and **c** 128/64. Clause accuracy of 3-SAT problems (blue) with clause-to-variable ratios, $m/n$, of **d** 32/32, **e** 40/24, and **f** 48/16. Cut values of MaxCut problems (purple) with $\mathbf{J}$ densities of **g** 20%, **h** 60%, and **i** 100%. MaxCut. Chip cut values (purple) are compared to cut values computed using an Apple M4 MacBook Pro (grey).

While cut values may be further optimized by fine tuning iteration count, learning rate, system non-linearity, and other hyperparameters, it is often the case that a sufficiently low energy state, i.e. local rather than global minimum, is acceptable as a practical solution for many applications such as deep learning. In fact, a global minimum can be detrimental in training neural networks and may lead to overfitting[37]. Nonetheless, there remains a trade-off between the quality of a solution and the time and energy expense required to reach such a solution. Here, with a clear positive, non-zero bias shown in the histograms, the chip shows its ability to effectively solve not only partially, but also fully connected problems.

**Time and energy to solution**

As a benchmark and figure of merit across machine architectures, one may quantify a time to solution (TTS) using the equation[1,9,11,25]

$$\mathrm{TTS}(P) = t_a \cdot \frac{\ln(1-P)}{\ln[1-P_a(t_a)]}, \tag{3}$$

where $t_a$ is the annealing time, $P$ is the desired probability (99%) of obtaining an optimal solution, and $P_a(t_a)$ is the machine's probability of providing an optimal solution given an annealing time of $t_a$. Since all NP problems may be reduced to a 3-SAT problem formulation[38,39] with a clear optimal solution (i.e. zero clause errors), we define the TTS based on the machine's performance on solving 3-SAT problems. Based on Figs. 4d – f, the probability, $P_a(t_a)$, of obtaining 100% clause accuracy is 35.4%, 20.8%, and 2.0% for clause-to-variable ratios of $m/n$ = 32/32, 40/24, and 48/16, respectively. Additionally, we define annealing time as the time required for all spins to bifurcate to 90% of their final value (see Fig. 3c), where, based on Extended Data Fig. 7d, the mean bifurcation time corresponds to approximately 11 round-trip iterations, regardless of clause-to-variable ratio. Given a round-trip time of 64 ns, this equates to an annealing time of $t_a \approx 700$ ns. Therefore, TTS(99%) is 7.4 $\mu$s, 13.8 $\mu$s, and 159.2 $\mu$s for $m/n$ = 32/32, 40/24, and 48/16, respectively. Similarly, given a dynamic chip power consumption of 400 mW, the energy to solution (ETS) is calculated as

$$\mathrm{ETS}(P) = \mathrm{TTS}(P) \times 400\ mW. \tag{4}$$

Thus, ETS(99%) is 2.9 $\mu$J, 5.5 $\mu$J, and 63.7 $\mu$J for $m/n$ = 32/32, 40/24, and 48/16, respectively.

## Discussion and conclusion

In this work, a compact all-to-all Ising solver is demonstrated that utilizes a low-energy mixed-signal CMOS integrated circuit working in concert with a low-cost, low-loss, high-bandwidth optical delay line. Together, these enable low energy and fast optimization while simultaneously enabling a high degree of scalability without the use of any benchtop devices within the loop. The mixed-signal integrated circuit, designed using TSMC's 65nm CMOS process, enables fast, energy-efficient analog current-mode multiplication and addition where high-speed digital memory is distributed throughout the chip for reconfigurability. Co-location of analog and digital blocks enables a low-power miniature implementation of the high-speed forward path of the Ising solver. More details regarding the analog signal processing and distributed digital memory architecture are included in Extended Data Figs. 1 and 2, respectively, and the Methods section. Record time and energy to solutions were demonstrated using the Ising solver in this work. These metrics may be further improved by moving to smaller fabrication technology nodes to improve the energy efficiency and compute speeds of the chip.

The system timing alignment is essential for correct and effective operation of the proposed Ising solver. The circuitry used to create the necessary timing waveforms for the chip is shown in Extended Data Fig. 3 with corresponding waveforms plotted in Extended Data Fig. 4. The 1:64 interleaver samples the output of the optical delay line, which serves as an analog memory for the spins as the system evolves. With correct loop delay, the system will sustain spin patterns as initially set by the $\mathbf{h}$ vector after the $\mathbf{h}$ path is disabled, as shown in Extended Data Figs. 5a – c. More details on the system time synchronization are included in the Methods section. To study the effect of delay variations/mismatches within the loop, the system timing alignment and sensitivity were measured, indicating the solution quality remains high over delay variations ranging from -50 ps $\leq \Delta t \leq$ 100 ps, making the system robust against delay variations (i.e. 150 ps of delay corresponds to about 3 cm of fiber). Additional details on the system timing and sensitivity are included in the Methods section and Extended Data Figs. 5d – e.

To characterize the nonlinearity of the system, the chip output was measured while the input applied to the chip was increased. A test $\mathbf{J}$ matrix was loaded into the chip memory to generate an output alternating ramp waveform. As shown in Extended Data Fig. 6, the chip output eventually saturates for large input voltages where the resulting nonlinear function is symmetric for both positive and negative spin values. More details on the nonlinearity characterization are included in the Methods section. The effect of other types of nonlinearities, including optical nonlinearities (i.e. laser or modulator), on the solution quality, the time to solution, and energy to solution may be explored in future implementations.

To determine the annealing time of the system, the 3-SAT problems were simulated on a model of the mixed-signal circuitry. Average annealing/bifurcation time is quantified as 11 round trips, or 700 ns, as shown in Extended Data Fig. 7. The annealing time is independent of the clause-to-variable ratio, and fast annealing times (on the order of 100's nanoseconds) are possible due to the high-bandwidth current-mode and optical signaling. Given the high bandwidth of the modulator and photodiode (i.e. larger than 10 GHz in the current implementation), the annealing time can be further reduced by increasing the clock frequency, which could result in lower energy and time to solution, especially if smaller fabrication technologies nodes are used.

While the Ising machine has been demonstrated on 64-spin problems, the architecture of the system is scalable to higher spin counts while remaining fully connected and maintaining a small footprint. Two scalability approaches are viable: (1) The on-chip memory of a single chip may be increased to support a larger J matrix, or (2) multi-chip tiling may be used where chips are co-packaged and memory is allocated across the entire multi-chip mosaic. This is further discussed in the Methods section and Extended Data Fig. 8.

The implemented Ising machine in this work is compared to other state-of-the-art all-to-all Ising solvers in Extended Data Table 1. Benchmarked on NP-complete 3-SAT problems, the proposed Ising chip achieves TTS as low as 7.4 $\mu$s, and corresponding ETS of 2.9 $\mu$J. The measurement setup and list of all components and devices used for characterization and measurement of the system are included in Extended Data Fig. 9.

The combination of a reliable low-energy mixed-signal CMOS chip and a compact low-loss high-bandwidth optical delay line realizes a low-cost highly energy-efficient and high-speed optimization platform in a small footprint without any benchtop devices or instruments in the loop. By utilizing high-bandwidth optical devices and current-mode signaling, the implemented system achieves annealing times on the order of 100's of nanoseconds, which is approximately three orders-of-magnitude lower than the state-of-the-art all-to-all Ising solvers. This results in more than an order of magnitude improvement in TTS and ETS. Despite excellent performance, many state-of-the art Ising solvers still heavily rely on benchtop devices such as FPGA boards, oscilloscopes, computers, arbitrary waveform generators, and/or MCUs to close the loop and perform spin couplings. Often, the power consumption and latency of these devices are omitted in the overall energy efficiency. When considered, these components limit the practicality of such systems, especially for lower-power, edge computing applications. Here, the CMOS integrated circuit allows for extreme improvement in time and energy, eliminating the need for any such benchtop devices or instruments. Additionally, by using phase-insensitive optical pulses and current-mode signaling, the proposed architecture obviates the need for complex locking mechanisms that are often required for phase-sensitive systems. Furthermore, amplitude-only optical processing does not necessitate high-Q or high-finesse resonators/cavities in the loop that fundamentally lead to large time and energy to solution[7,9]. Resonator Q also affects TTS and ETS of CMOS and $VO_2$ oscillator-based machines[13] which are reliant on transient interactions with time constants set by the quality factor of the oscillator.

With recent advancements in monolithic photonic-electronic processes, such as those offered by GlobalFoundries[40], the proposed photonic-assisted Ising architecture has potential to further reduce power and energy consumption by integrating photonic devices monolithically with mature CMOS devices. Recent state-of-the-art monolithic electro-optical transceivers have shown bandwidth densities of 4 Tb/s/mm and energy efficiencies less than 100 fJ/bit[18-22]. In this way, the photodiode, TIA, modulator, and other photonic and electronic devices required for sensing and data transmission may all be monolithically integrated with the rest of the system. With hybrid integration of a low-cost laser, the proposed architecture enables the low-cost and large-scale deployment of a low-energy, fast, and compact Ising solver for future compute and AI systems.

## Methods

### Optimization algorithm

The chip implements gradient descent to find a minimum energy state. In this case, as the system evolves, spin values are perturbed by the derivative of the Hamiltonian. The spin value is updated as

$$\sigma_i^{t+1} = F_{NL}\left(\beta\sigma_i^t - \alpha\partial_{\boldsymbol{\sigma}}\big(H(\boldsymbol{\sigma})\big)_i\right), \tag{5}$$

where $\beta$ represents the gain of the feedforward path, $\alpha$ represents the coefficient of the derivative of the Hamiltonian, and $F_{NL}(\cdot)$ represents the nonlinear function of the saturable output driver. Given the Hamiltonian of the Ising system is $H = -\boldsymbol{\sigma}^T \mathbf{J}\boldsymbol{\sigma} - \mathbf{h}^T\boldsymbol{\sigma}$, as shown in Eqn. (1), the derivative of the Hamiltonian can be calculated as

$$\alpha\partial_{\boldsymbol{\sigma}}\big(H(\boldsymbol{\sigma})\big) = -2\mathbf{J}\boldsymbol{\sigma} - \mathbf{h}. \tag{6}$$

Therefore, each spin value is updated as

$$\sigma_i^{t+1} = F_{NL}(\beta\sigma_i^t + \alpha(2\mathbf{J}\boldsymbol{\sigma} + \mathbf{h})_i). \tag{7}$$

In this case, the gains of the $\mathbf{J}\boldsymbol{\sigma}$ and $\mathbf{h}$ paths in Fig. 2a are $2\alpha$ and $\alpha$, respectively. The ratio $\alpha/\beta$ is defined as the learning rate hyperparameter.

To implement Eqn. (7) on chip, the feedforward ($\beta\sigma_i^t$) path, matrix-vector multiplication ($\mathbf{J}\boldsymbol{\sigma}$) path, and bias ($\mathbf{h}$ vector) path are summed on-chip in the current domain. In this equation, $\sigma_i^t$ represents the present spin input to the chip and $F_{NL}(\cdot)$ represents the nonlinear characteristics of the on-chip output driver producing the updated spin value, $\sigma_i^{t+1}$. The parameter $\alpha$ can be adjusted by changing the current bias of the multipliers and adders (Extended Data Figs. 1c – d). The following sections describe additional details of how this equation is realized via mixed-signal circuitry.

**Analog circuit architecture**

The diagram of the analog signal processing path within the chip is presented in Extended Data Fig. 1a, where current-mode multiplication and addition allow for operation at 1 GHz for a total round-trip iteration time of 64 ns for 64 spins. The sub-blocks of the analog path are shown in Extended Data Figs. 1b – f. The analog signal processing is utilized for two main tasks: (1) sensing and storing the incoming spin values, $\boldsymbol{\sigma}$, and (2) performing the MAC operations required to update each spin. For the first task, all 64 spins are sampled and stored in analog form using a 1:64 interleaver, implemented as two back-to-back 1:8 interleaver stages. Source followers (yellow) are used as buffers to drive the routing capacitance with adequate bandwidth while suppressing charge sharing between sampling nodes. The first buffer stage is designed using a PMOS source follower, while the second buffer stage is an NMOS source follower with an active load. Switches are implemented using CMOS transmission gates and clocked using the circuitry and waveforms shown in Extended Data Figs. 3 and 4, respectively. The interleaver consumes approximately 75 mW based on post-layout simulations.

After sampling $\boldsymbol{\sigma}$ via the 1:64 interleaver, the second task of the analog circuitry is to perform the MAC operations required for performing the all-to-all $\mathbf{J}\boldsymbol{\sigma}$ coupling. The 4-bit programmable coupling between the spins is implemented using a current-mode multiplier, where 3 bits are used for amplitude control and 1 bit is used for polarity (sign) control, as shown in Extended Data Fig. 1c.

The frequency response of the multiplier is shown in Extended Data Fig. 1e. To ensure settling within one clock period (1 ns) and to minimize inter-symbol interference, gain flatness is maintained within 0.1 dB from DC to 1 GHz for all codes. To match the delays between the pulses in the $\mathbf{J}\boldsymbol{\sigma}$ and $\mathbf{h}$ vector paths, an identical multiplier with constant input is used to implement $\mathbf{h}$ (green in Extended Data Fig. 1a).

Following the multiplication of $\mathrm{J}_{i,j} \cdot \sigma_j$, accumulation is performed using a 64:1 analog summer, implemented as three cascaded stages of 4:1 current-mode adders. The schematic of a single 4:1 adder is shown in Extended Data Fig. 1d, where the differential input voltages of the adder are converted to currents using a source-degenerated differential pair. The resulting currents are summed at a shared node, and the aggregate current is converted back to a voltage via a transimpedance resistor, serving as the input to the subsequent stage.

A clocked feedforward path samples the incoming spin, $\sigma_i$, and amplifies it by a gain of $\beta$ (orange in Extended Data Fig. 1a). The outputs of the feedforward $\mathbf{J}\boldsymbol{\sigma}$ and $\mathbf{h}$ vector paths are added using transconductance stages as shown in Extended Data Fig. 1f. The resulting current is routed to the input of the on-chip output driver and also made available on pads (“External sum”) to support scaling via tiling.

Pulsed spins are generated by a pulsed pre-driver, followed by a nonlinear amplification stage, as shown in Extended Data Fig. 1b. For high-speed operation, the pre-driver is pulsed by switching the tail current of a differential pair, while the subsequent cascode amplifier provides additional gain. The nonlinearity is primarily realized by increasing the tail current of the second cascode stage, producing the desired bistable transfer characteristic of $F_{NL}(\cdot)$ as shown in Extended Data Fig. 6.

**On-chip memory architecture**

Dense CMOS integration allows for high-speed digital memory to be distributed throughout the chip, where memory units are placed next to each analog block. The memory architecture of the Ising chip is shown in Extended Data Fig. 2a with digital sub-blocks depicted in Extended Data Figs. 2b – d. A total of 16,384 bits are responsible for storing the entire 64 x 64 $\mathbf{J}$ matrix (blue), and 256 bits are responsible for storing the 64 elements of the $\mathbf{h}$ vector (green). Each multiplier, including the one used for the $\mathbf{h}$ vector, has a local memory of 64 x 4 bits corresponding to a 4-bit coupling weight for each of the 64 spins. Multiplexers are used to access the weights stored in each local memory bank, allowing for high-speed memory access. The select lines of the multiplexers are driven using the address bits produced by the frequency divider, where the outputs of the digital divider serve as the address bits for the memory, as shown in Extended Data Fig. 3a.

As depicted in Extended Data Fig. 2b, each 64 x 4-bit memory bank is composed of four 64 x 1-bit memory cells, where each cell is responsible for 1 bit of the total 4-bit coupling resolution. The 64 x 1-bit memory cell is formed using a series of 64 D flip-flops, as shown in Extended Data Fig. 2c. The 64 outputs of the D flip-flop memory bank are accessed via a 64:1 multiplexer, which is composed of 6 layers of 2:1 multiplexer, as shown in Extended Data Fig. 2d. The 2:1 multiplexer is designed using CMOS transmission gates, and an output inverter is used to regenerate the signal.

A chip select line enables loading the on-chip memory via serial peripheral interface (SPI) with data and clock signals provided on shift register (SR) data and SR clock, respectively. In this design, 64 x 4 bits are allocated to the $\mathbf{h}$ vector (green), and 64 x 64 x 4 bits are allocated to the $\mathbf{J}$ matrix multiplier (blue) for a total on-chip memory of 16,640 bits. The $\mathbf{J}$ matrix is partitioned into 4 blocks: $k$ = 0, 1, 2, and 3, where the first block is responsible for MAC operations 1 through 16 (MAC<1:16>), the second block for MAC operations 17 through 32 (MAC<17:32>), the third block for MAC operations 33 through 48 (MAC<33:48>), and the fourth block for MAC operations 49 through 64 (MAC<49:64>). The shift registers within each block are serially connected, where the last shift register output drives a pad to allow multiple chips to be daisy-chained for scaling to higher spin counts in the future.

**On-chip clock generation**

The on-chip clock generator takes a 1 GHz input reference and divides it by a factor of 256 using 8 cascaded divide-by-two frequency dividers, as shown in Extended Data Fig. 3a. The first 6 outputs of the divider chain are used as the address for the on-chip memory. The ÷128 and ÷256 outputs are used to synchronize the data capture with the output of the chip and can be used to time-gate multi-chip solutions for scaling to higher spins counts as shown in Extended Data Fig. 8. To create the appropriate clock phases for the on-chip 1:64 interleaver, the divider outputs are further processed by OR and NOR gates, as shown in Extended Data Figure 3b. A 3-input NOR gate creates $\varphi_P$ pulse which has a 1 ns width and 8 ns period (green). A cascade of a 3-input OR gate with a 2-input NOR gate is used to generate $\varphi_Q$, which has a 4 ns pulse width and 64 ns period (blue). Finally, a 2-input NOR gate is used to generate $\varphi_R$, a pulse train with 1 ns pulse width and 4 ns period (purple). Next, 3 sets of 8 cascaded D flip-flops are used to create delayed versions of each of these signals, as shown in Extended Data Fig. 3c. The delayed versions of $\varphi_P$, corresponding to $\varphi_1$ through $\varphi_8$, are used to clock the first 1:8 interleaving stage, while the circuit shown in Extended Data Fig. 3d is used to create the clock phases ($\varphi_{1,1}$ through $\varphi_{8,8}$) for the second interleaving stage. The complete set of phases used to clock the 1:64 interleaver is shown in Extended Data Fig. 4.

**System timing alignment**

A calibration vector was used to verify the 64 ns round-trip time and ensure pulses going around the loop are sampled correctly (i.e. in a synchronized way) by the chip. Since the chip processes 64 spins at 1 GHz, the output of the chip must be delayed 64 ns before it can be sampled again. Coarse adjustments were made to the optical delay by cutting and splicing an optical fiber to approximately 12.8 m. After splicing, the delay was adjusted in analog increments using the tunable optical delay line. To verify timing alignment, the chip memory was loaded with a calibration $\mathbf{h}$ vector consisting of alternating −1, +1 values for the first 16 spins, and +1 values for the remaining 48 spins, as shown in Extended Data Fig. 5a. Next, as depicted in Extended Data Fig. 5b, with the $\mathbf{J}$ matrix disabled, the feedforward gain, $\beta$, was increased such that the amplified version of the calibration $\mathbf{h}$ vector appeared on the output, confirming that the chip was correctly sampling the input optical pulse train. Finally, while the chip output was monitored, the $\mathbf{h}$ vector was disabled, and the pattern originally set by $\mathbf{h}$ was sustained within the delay line, confirming that the optical delay line could correctly store the analog spins (Extended Data Fig. 5c).

### System timing sensitivity

The timing sensitivity of the system was characterized using the setup shown in Extended Data Fig. 5d (left), where the delay offset of the tunable optical delay line was adjusted by $\Delta t$ within the range of -180 ps $\leq \Delta t \leq$ 180 ps in 10 ps increments. At each delay offset, A MaxCut problem was solved 10 times with each cut value shown in Extended Data Fig. 5d (right). The cut value was determined by finding the peak of the waveform within 1 ns increments to determine the polarity of each spin. To demonstrate how the output waveform changes based on the delay offset, three transient waveforms were selected at delay offsets of -180 ps, 0 ps, and 180 ps, as shown in green, blue, and purple, respectively. For these delay offsets, the transient chip output, Vout, is plotted in Extended Data Fig 5e, indicating that there is approximately a 150 ps range (between -50 ps and 100 ps) where the chip continues to provide an adequate solution, verifying that the system is robust against delay variations (i.e. 150 ps of delay corresponds to about 3 cm of fiber).

### Nonlinearity measurements

To demonstrate how the system saturates, the loop was opened, and a sample $\mathbf{J}$ matrix was loaded into the chip where the first 32 spins are nulled, and the last 32 spins exhibit an alternating ramp waveform, as shown in Extended Data Fig. 6a. A DC input bias was applied to the chip such that each sampling capacitor holds the same value. For small input voltages, the ramp waveform is linear. However, as the differential input to the chip is increased, the ramp waveform becomes distorted as the loop gain drops for larger amplitudes, as shown in Extended Data Figs. 6b – c. The corresponding nonlinear transfer function is shown in Extended Data Fig. 6d.

### Mean bifurcation time

To quantify time to solution, a model of the chip was simulated using the same 3-SAT problems run on the hardware. During simulation, the evolution of the spins was captured as the feedforward path was enabled. An example evolution of $\boldsymbol{\sigma}$ is provided in Fig. 3c. The bifurcation/annealing time is quantified as the number of round-trip iterations for all spins to settle to 90% of their final value from the moment $\beta$ is activated. In this case, similar to the system measurements, all 300 unique 3-SAT problems were run 200 times in simulation with histograms of the settling times shown in Extended Data Fig. 7a – c. The corresponding box charts are shown in Extended Data Fig. 7d, where the mean bifurcation time is approximately 10.9 round-trip iterations, regardless of the clause-to-variable ratio. As a result, the average annealing time, for a round-trip time of 64 ns (at 1 GHz clock), can be approximated as 700 ns.

### System scalability

The current 64-spin chip stores 4,096 4-bit coupling weights (for a 64 x 64 $\mathbf{J}$ matrix), for a total memory size of 16,384 bits (excluding the $\mathbf{h}$ vector). If the number of spins increases by a factor of $x$, the size of the $\mathbf{J}$ matrix scales by a factor of $x^2$. Therefore, for example, scaling to 256 spins requires a total on-chip $\mathbf{J}$ matrix memory of only 262,144 bits. The digital memory architecture of the proposed Ising chip allows this to be performed with minimal area overhead and scales favorably with smaller technology nodes. Furthermore, ultradense memory architectures have been reported demonstrating bit densities exceeding 29 Gb/mm$^2$ and show promise for extremely large, highly connected systems[41]. Alternatively, to expand the spin count, multi-chip tiling may be leveraged to allocate memory across multiple chips.

As an example, a roadmap for scaling to 256 spins via tiling is presented in Extended Data Fig. 8. In this case, 16 chips may form a mosaic to store a total of 256 x 256 $\mathbf{J}$ matrix values, where each individual chip stores 1/16$^{th}$ of the 256 x 256 $\mathbf{J}$ matrix. A tile of 4 chips is created by adding the output of each chip in the current domain via the external summation pads shown in Extended Data Fig. 1a and using a single output driver among the four chips. Four clock phases, $\Phi_1$ through $\Phi_4$, may be derived from the ÷128 and ÷256 chip outputs (Extended Data Fig. 3a) via the logic circuits shown in Extended Data Fig. 8c to produce the waveforms shown in Extended Data Fig. 8d. In addition to timing the output drivers, the clock phases are used to time-gate the interleavers of each chip such that each tile holds a sample of all 256 spins, as labeled next to each interleaver in Extended Data Fig. 8a. In this way, the first tile is responsible for $\mathrm{J}_{1,1}$ through $\mathrm{J}_{64,256}$, the second tile for $\mathrm{J}_{65,1}$ through $\mathrm{J}_{128,256}$, the third tile for $\mathrm{J}_{129,1}$ through $\mathrm{J}_{192,256}$, and the fourth tile for $\mathrm{J}_{193,1}$ through $\mathrm{J}_{256,256}$, as shown in Extended Data Fig. 8b. The density of mature CMOS processes allows scalability to be performed with minimal area overhead, while the low loss of the optical fiber and small footprint of a fiber spool[42] facilitate efficient data transfer between multiple chips. In this way, the implemented Ising machine may be scaled to a system with a larger number of spins while still occupying a small footprint.

## Data availability

The data that support the findings of this study are available from the corresponding author upon reasonable request.

## Code availability

The code used in the present work is available from the corresponding author upon reasonable request.

## Acknowledgement

This work was funded by Defense Advanced Research Project Agency QuICC program (FA8750-23-C-0539). The authors would like to thank Jaeung Ko for the chip micrograph.

## Author contributions

H.L., Z.Y., S.P. and F.A. conceived the project idea. H.L. and Z.Y. designed, simulated, and performed layout of the chip. H.L., Z.Y., and B.E. made the model of the system, and M.A. and B.E. perform behavioral simulations of the chip. H.L. designed the PCB. H.L., Z.Y., M.A., B.E., and P.C. formulated the Ising problems. Y. J. and N. C. contributed to the system design. H.L., Z.Y., and M.A. conducted the measurements. All authors contributed to writing the manuscript led by H.L. F.A. directed and supervised the project.

## Competing interests

The authors declare no competing interests.

## Additional information

**Supplementary information** The online version contains supplementary material available at [link].

**Correspondence** and request for materials should be addressed to Firooz Aflatouni.

**Peer review information** TBD

**Reprints and permissions information** is available at www.nature.com/reprints.

## References

[1] Yamamoto, K. et al. STATICA: A 512-Spin 0.25M-Weight Annealing Processor With an All-Spin-Updates-at-Once Architecture for Combinatorial Optimization With Complete Spin–Spin Interactions. *IEEE J. Solid-State Circuits* **56**, 165–178 (2021).
DOI: https://doi.org/10.1109/JSSC.2020.3027702

[2] Su, Y., Mu, J., Kim, H. & Kim, B. A Scalable CMOS Ising Computer Featuring Sparse and Reconfigurable Spin Interconnects for Solving Combinatorial Optimization Problems. *IEEE J. Solid-State Circuits* **57**, 858–868 (2022).
DOI: https://doi.org/10.1109/JSSC.2022.3142896

[3] Ahmed, I., Chiu, P.-W., Moy, W. & Kim, C. H. A Probabilistic Compute Fabric Based on Coupled Ring Oscillators for Solving Combinatorial Optimization Problems. *IEEE J. Solid-State Circuits* **56**, 2870–2880 (2021).
DOI: https://doi.org/10.1109/JSSC.2021.3062821

[4] Lo, H., Moy, W., Yu, H., Sapatnekar, S. & Kim, C. H. An Ising solver chip based on coupled ring oscillators with a 48-node all-to-all connected array architecture. *Nat. Electron.* **6**, 771–778 (2023).
DOI: https://doi.org/10.1038/s41928-023-01021-y

[5] Cılasun, H. et al. A coupled-oscillator-based Ising chip for combinatorial optimization. *Nat. Electron.* **8**, 537–546 (2025).
DOI: https://doi.org/10.1038/s41928-025-01393-3

[6] McMahon, P. L. et al. A fully programmable 100-spin coherent Ising machine with all-to-all connections. *Science* **354**, 614–617 (2016).
DOI: https://doi.org/10.1126/science.aah5178

[7] Honjo, T. et al. 100,000-spin coherent Ising machine. *Sci. Adv.* **7**, eabh0952 (2021).
DOI: https://doi.org/10.1126/sciadv.abh0952

[8] Wei, H. et al. A versatile coherent Ising computing platform. *Light Sci. Appl.* **15**, 74 (2026).
DOI: https://doi.org/10.1038/s41377-025-02178-1

[9] Jin, Y. et al. A Kerr soliton Ising machine for combinatorial optimization problems. Preprint at https://arxiv.org/abs/2508.00810 (2025).
DOI: https://doi.org/10.48550/arXiv.2508.00810

[10] Hu, Z. et al. Programmable optoelectronic Ising machine for optimization of real-world problems. *Light Sci. Appl.* **15**, 6 (2026).
DOI: https://doi.org/10.1038/s41377-025-02100-9

[11] Al-Kayed, N. et al. Programmable 200 GOPS Hopfield-inspired photonic Ising machine. *Nature* **648**, 576–584 (2025).
DOI: https://doi.org/10.1038/s41586-025-09838-7

[12] Wu, B. et al. A monolithically integrated optical Ising machine. *Nat. Commun.* **16**, 4296 (2025).
DOI: https://doi.org/10.1038/s41467-025-59537-0

[13] Maher, O. et al. A CMOS-compatible oscillation-based $VO_2$ Ising machine solver. *Nat. Commun.* **15**, 3334 (2024).
DOI: https://doi.org/10.1038/s41467-024-47642-5

[14] Si, J. et al. Energy-efficient superparamagnetic Ising machine and its application to traveling salesman problems. *Nat. Commun.* **15**, 3457 (2024).
DOI: https://doi.org/10.1038/s41467-024-47818-z

[15] Whitehead, W., Nelson, Z., Camsari, K. Y. & Theogarajan, L. CMOS-compatible Ising and Potts annealing using single-photon avalanche diodes. *Nat. Electron.* **6**, 1009–1019 (2023).
DOI: https://doi.org/10.1038/s41928-023-01065-0

[16] Sipser, M. *Introduction to the Theory of Computation*. 3rd ed. Cengage Learning. Boston, MA (2013).

[17] Fortnow, L. The status of the P versus NP problem. *Commun. ACM* **52**, 78–86 (2009).
DOI: https://doi.org/10.1145/1562164.1562186

[18] Daudlin, S., Rizzo, A., Lee, S. et al. Three-dimensional photonic integration for ultra-low-energy, high-bandwidth interchip data links. *Nat. Photon.* **19**, 502–509 (2025).
DOI: https://doi.org/10.1038/s41566-025-01633-0

[19] Rizzo, A., Novick, A., Gopal, V. et al. Massively scalable Kerr comb-driven silicon photonic link. *Nat. Photon.* **17**, 781–790 (2023).
DOI: https://doi.org/10.1038/s41566-023-01244-7

[20] Pirmoradi, A. et al. Sing-Chip 1.024 Tb/s Optical Receiver for High-Speed Optical Links. Preprint at https://arxiv.org/abs/2601.07709 (2026).
DOI: https://doi.org/10.48550/arXiv.2601.07709

[21] Omirzakhov, K. & Aflatouni, F. 12.1 Monolithically Integrated Sub-63 fJ/b 8-Channel 256Gb/s Optical Transmitter with Autonomous Wavelength Locking in 45nm CMOS SOI. In *2024 IEEE International Solid-State Circuits Conference (ISSCC)*, 218–220 (IEEE, 2024).
DOI: https://doi.org/10.1109/ISSCC49657.2024.10454519

[22] Shoobi, A., Omirzakhov, K., Yu, Z., Pirmoradi, A., & Aflatouni, F. 14.2 An 8λ×38Gb/s/λ 106fJ/b Optical WDM Transmitter in 45nm CMOS SOI. In *2026 IEEE International Solid-State Circuits Conference (ISSCC)*, 242–244, (IEEE, 2026).
DOI: https://doi.org/10.1109/ISSCC49663.2026.11409254

[23] Yu, G. (2013). *Industrial Applications of Combinatorial Optimization*. Springer. New York, NY (2013).
DOI: https://doi.org/10.1007/978-1-4757-2876-7

[24] Rabaey, J. M., Chandrakasan, A. & Nikolić, B. *Digital Integrated Circuits*. 2nd ed. Pearson Education, Inc. (2003).

[25] Garg, C. & Salahuddin, S. Efficient optimization accelerator framework for multi-state spin Ising problems. *Nat. Commun.* **16**, 9601 (2025).
DOI: https://doi.org/10.1038/s41467-025-64625-2

[26] Nikhar, S., Kannan, S., Aadit, N. A., Chowdhury, S. & Camsari, K. Y. All-to-all reconfigurability with sparse and higher-order Ising machines. *Nat. Commun.* **15**, 8977 (2024).
DOI: https://doi.org/10.1038/s41467-024-53270-w

[27] Onizawa, N. & Hanyu, T. GPU-accelerated simulated annealing based on p-bits with real-world device-variability modeling. *Sci. Rep.* **15**, 6118 (2025).
DOI: https://doi.org/10.1038/s41598-025-90520-3

[28] Markov, I. L. Limits on fundamental limits to computation. *Nature* **512**, 147–154 (2014).
DOI: https://doi.org/10.1038/nature13570

[29] Ising, E. Beitrag zur Theorie des Ferromagnetismus. *Z. Physik* **31**, 253–258 (1925).
DOI: https://doi.org/10.1007/BF02980577

[30] Lucas, A. Ising formulations of many NP problems. *Front. Physics* **2** (2014).
DOI: https://doi.org/10.3389/fphy.2014.00005

[31] Kirkpatrick, S., Gelatt, C. D. & Vecchi, M. P. Optimization by Simulated Annealing. *Science* **220**, 617–680 (1983)
DOI: https://doi.org/10.1126/science.220.4598.671

[32] Feldman, M. R., Esener, S. C., Guest, C. C. & Lee, S. H. Comparison between optical and electrical interconnects based on power and speed considerations. *Appl. Opt.* **27**, 1742–1751 (1988).
DOI: https://doi.org/10.1364/AO.27.001742

[33] Petrovich, M. et al. Broadband optical fibre with an attenuation lower than 0.1 decibel per kilometre. *Nat. Photon.* **19**, 1203–1208 (2025).
DOI: https://doi.org/10.1038/s41566-025-01747-5

[34] Habib, M. S. & Correa, R. A. Hollow-core breakthrough. *Nat. Photon.* **19**, 1160–1161 (2025).
DOI: https://doi.org/10.1038/s41566-025-01774-2

[35] Nüßlein, J., Gabor, T., Linnhoff-Popien, C. & Feld, S. Algorithmic QUBO formulations for *k*-SAT and hamiltonian cycles. In *Proc. Genetic and Evolutionary Computation Conference (GECCO)*, 2240–2246 (2022).
DOI: https://doi.org/10.1145/3520304.3533952

[36] MacBook Pro (14-inch, M4, 2024) – Tech Specs. https://support.apple.com/en-us/121552 (2024).
DOI: https://support.apple.com/en-us/121552

[37] Choromanska, A., Henaff, M., Mathieu, M., Arous, G. B. & LeCun, Y. The loss surfaces of multilayer networks. In *Proceedings of the 18th International Conference on Artificial Intelligence and Statistics* 192–204 (PMLR, 2015).
DOI: https://proceedings.mlr.press/v38/choromanska15.pdf

[38] Cook, S. A. The complexity of theorem-proving procedures. In *Proc. 3rd Annual Symposium on Theory of Computing (STOC)*, 151–158 (1971).
DOI: https://doi.org/10.1145/800157.805047

[39] Levin, L. A. Universal Sequential Search Problems. *Problems of Information Transmission* **9**, 265–266 (1973).
Link: https://lance.fortnow.com/papers/files/Levin%20Universal.pdf

[40] Rakowski, M. et al. 45nm CMOS – Silicon Photonics Monolithic Technology (45CLO) for Next-Generation, Low Power and High Speed Optical Interconnects. In *2020 Optical Fiber Communications Conference and Exhibition (OFC)*, 1–3 (IEEE, 2020).
DOI: https://doi.org/10.1364/OFC.2020.T3H.3

[41] Yanagidaira, K. et al. A 29-Gb/mm$^2$ 1-Tb 3-b/Cell 3-D Flash Memory With CMOS Direct Bonded Array (CBA) Technology. *IEEE J. Solid-State Circuits* **61**, 237–249 (2026).
DOI: https://doi.org/10.1109/JSSC.2025.3617579

[42] Shang, K. et al. Ultra-small interferometric fiber optic gyroscope with an integrated optical chip. *Chin. Opt. Lett.* **20**, 040601 (2022).
DOI: https://doi.org/10.3788/COL202220.040601

## Extended data figures and tables

| | [7] | [9] | [11] | [14] | [1] | | [5] | This work |
|---|---|---|---|---|---|---|---|---|
| **Technology** | Benchtop optics and electronics | Benchtop optics and electronics | TFLN modulators and benchtop electronics | SMTJ and benchtop electronics | 65 nm CMOS | | 65 nm CMOS | 65 nm CMOS |
| **Spin representation** | DOPO pulses | Kerr-solitons | TFLN modulated optical pulses | Nanomagnetic states | Digital | | Ring oscillator phases | **Phase incoherent optical pulses** |
| **Number of spins** | 100,000 | 256 | 256 (1) | 80 | 512 | | 59 | 64 |
| **Connectivity** | All-to-all | All-to-all | All-to-all | All-to-all | All-to-all | | All-to-all | **All-to-all** |
| **Weight resolution** | 1.5 bits | N/R | 3.3 ENOB (2) | N/A | 5 bits | | 4.8 bits | 4 bits |
| **Coupling mechanism** | FPGA + optics | Oscilloscope + computer + AWG | Oscilloscope + computer + AWG | MCU + comparator array + DAC | Digital logic | | Oscillator interaction | **Current-mode on CMOS chip** |
| **Bias field ($\mathrm{h}$)** | No | Yes | Yes | No | Yes | | No | **Yes** |
| **Annealing time** | 593 $\mu$s | 175 $\mu$s (3) | 14.08 s (4) | 100 $\mu$s/iteration | 130 $\mu$s | 480 $\mu$s | 100 $\mu$s | **700 ns** |
| **Time to solution** | N/R | 1 ms (3, 5, 10) | 14.08 s | 40 s (6) | 8.23 ms (7) | 1.5 ms (7) | N/R | **7.4 $\mu$s (8)**<br>**13.8 $\mu$s (9)**<br>**159.2 $\mu$s (10)** |
| **Energy to solution** | N/R | 500 $\mu$J (3, 5, 10, 11) | 62 J (12, 13) | 25.6 mJ (6) | 4.9 mJ (7, 13) | 885 $\mu$J (7, 13) | N/R | **2.9 $\mu$J (8)**<br>**5.5 $\mu$J (9)**<br>**63.7 $\mu$J (10)** |

(1) Higher spin counts supported for sparse connectivity
(2) For 106 GBaud operation
(3) Excludes time for feedback calculation on computer
(4) Per supplemental material, TTS is reduced TTS = $\tau_{ann} = t_a$ due to 100% success probability on number partitioning problems
(5) Estimated from figure
(6) For TSP70; includes data transfer time and 400,000 iterations for a good solution
(7) TTS(99%) / ETS(99%) for K2000 MaxCut benchmark
(8) TTS(99%) / ETS(99%) for 3-SAT with m/n = 32/32
(9) TTS(99%) / ETS(99%) for 3-SAT with m/n = 40/24
(10) TTS(99%) / ETS(99%) for 3-SAT with m/n = 48/16
(11) Only accounts for optical power; electrical power is not included
(12) Based on power reported in Table S5 for N = 1
(13) Calculated based on reported average power and time to solution

TFLN: thin-film lithium niobate, SMTJ: superparamagnetic tunnel junction, CMOS: complementary metal-oxide semiconductor, DOPO: degenerate optical parametric oscillator, N/A: not applicable, N/R: not reported, FPGA: field-programmable gate array, MCU: microcontroller unit, DAC: digital-to-analog converter, AWG: arbitrary waveform generator, TSP70: 70-city traveling salesman problem; TTS(99%) / ETS(99%) time/energy to solution with 99% probability of arriving at an optimal solution

**Extended Data Table 1 | Performance summary and comparison to other state-of-the-art all-to-all Ising machines**

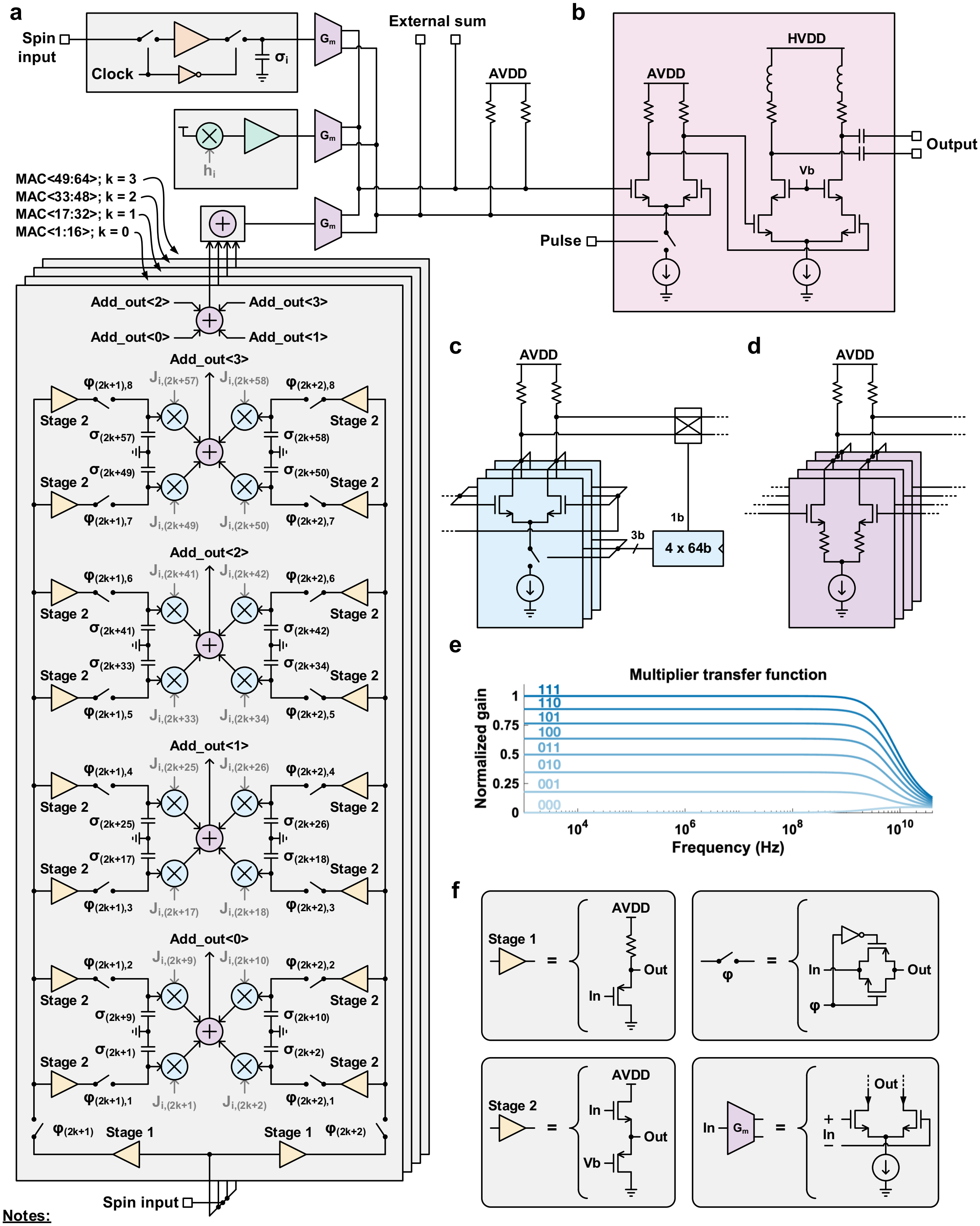

**Notes:**

- MAC<x:y> represents multiplication and accumulation (MAC) operations x through y where a complete $J\sigma$ matrix-vector multiplication requires 64 MAC operations: MAC<1:64>
- k is a variable to index $\sigma_j$; i.e. for the first memory block where k = 0, the bottom-most 4:1 MAC cell is responsible for $J_{i,1}\sigma_1 + J_{i,2}\sigma_2 + J_{i,9}\sigma_9 + J_{i,10}\sigma_{10}$

**Extended Data Fig. 1 | Analog signal path of the Ising chip. a** Core 64 x 64 $\mathbf{J}$ matrix multiplier with 2-stage interleaver (yellow), 4-bit multipliers (blue), and output adder tree (purple). The output driver (pink) amplifies the sum of the feedforward (orange), $\mathbf{J}\boldsymbol{\sigma}$, and $\mathbf{h}$ vector (green) paths. **b** Circuit diagram of the output driver. **c** Circuit diagram of the 4-bit multiplier. **d** Circuit diagram of each 4:1 adder. **e** Transfer function of the 4-bit multiplier (sign bit not shown). **f** Analog sub-blocks.

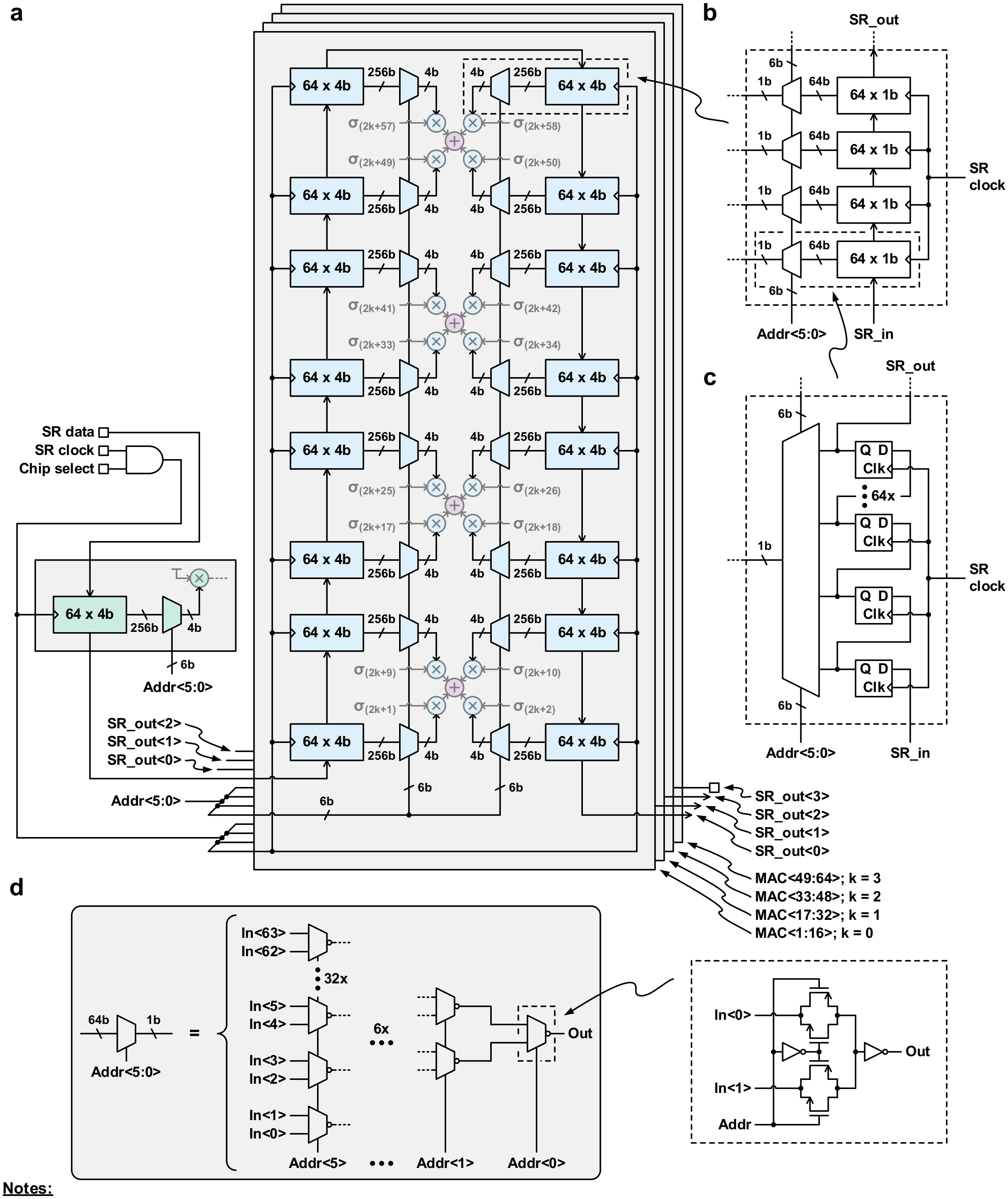

**Notes:**

- MAC<x:y> represents multiplication and accumulation (MAC) operations x through y where a complete $J\sigma$ matrix-vector multiplication requires 64 MAC operations: MAC<1:64>
- k is a variable to index $\sigma_j$; i.e. for the first memory block where k = 0, the bottom-most 4:1 MAC cell is responsible for $J_{i,1}\sigma_1 + J_{i,2}\sigma_2 + J_{i,9}\sigma_9 + J_{i,10}\sigma_{10}$

**Extended Data Fig. 2 | Memory architecture of the Ising chip. a** Each multiplier, including the **h** vector, has a local memory of 64 x 4 bits, corresponding to a 4-bit coupling weight for each of the 64 spins. The local memory is addressed using the address bits, Addr<5:0>, generated in Extended Data Fig. 3a. **b** Each 64 x 4-bit memory cell is composed of four 64 x 1-bit memory cells. **c** The 64 x 1-bit memory cells are composed of 64 serial static D flip-flops. **d** The local memory is accessed using a 64:1 multiplexer made from 6 layers of 2:1 multiplexers.

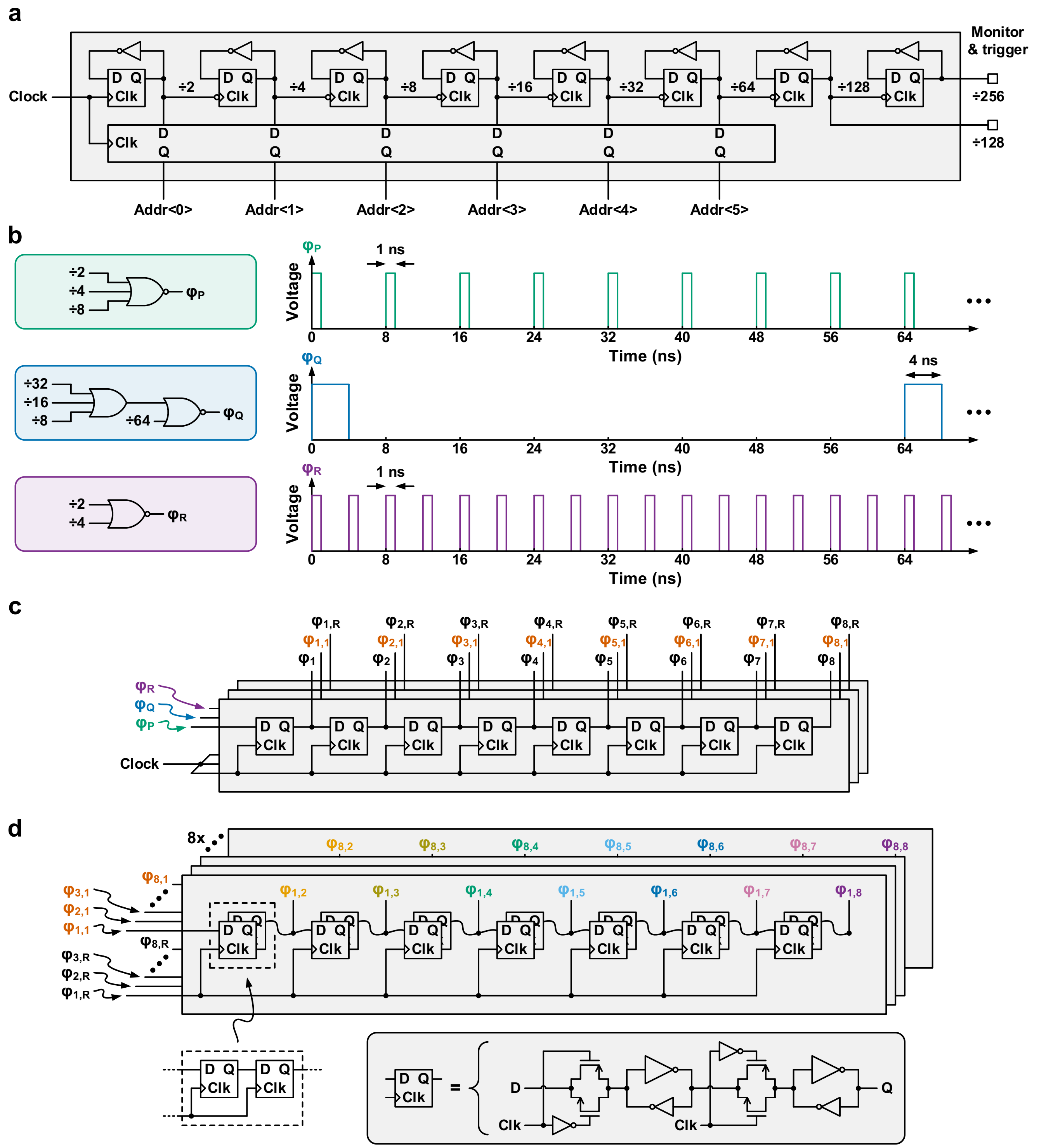

**Extended Data Fig. 3 | Clock generation circuitry. a** Frequency divider used to divide the 1 GHz input reference where the address for the memory, Addr<5:0>, is derived from the divider outputs. **b** OR and NOR gates are used to create $\varphi_P$, $\varphi_Q$, and $\varphi_R$. **c** 3 sets of 8 cascaded D flip-flops are used to generate delayed versions of these pulses. $\varphi_1$ through $\varphi_8$ are used to clock the first interleaving stage, while the circuit in **d** is used to create the clock phases, $\varphi_{1,1}$ through $\varphi_{8,8}$, for the second interleaving stage. The clocking waveforms are further shown in Extended Data Fig. 4.

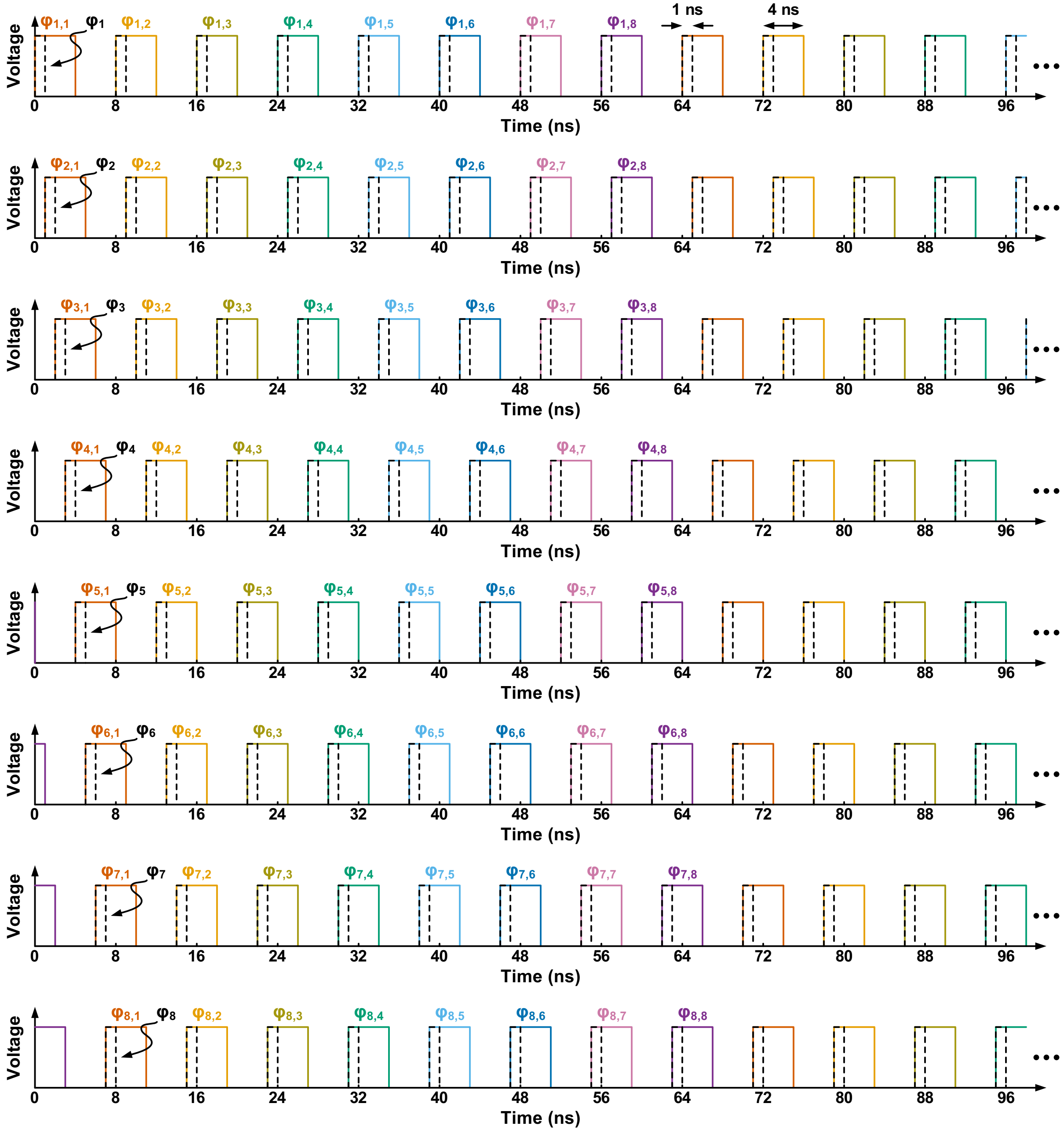

**Extended Data Fig. 4 | Clock waveforms for the 1:64 interleaver.** $\varphi_1$ through $\varphi_8$ (dashed) are used to clock the first 1:8 interleaver stage with a pulse width and period of 1 ns and 8 ns, respectively. $\varphi_{1,1}$ through $\varphi_{8,8}$ (solid) are used to clock the second 1:8 interleaver stage with a pulse width and period of 4 ns and 64 ns, respectively.

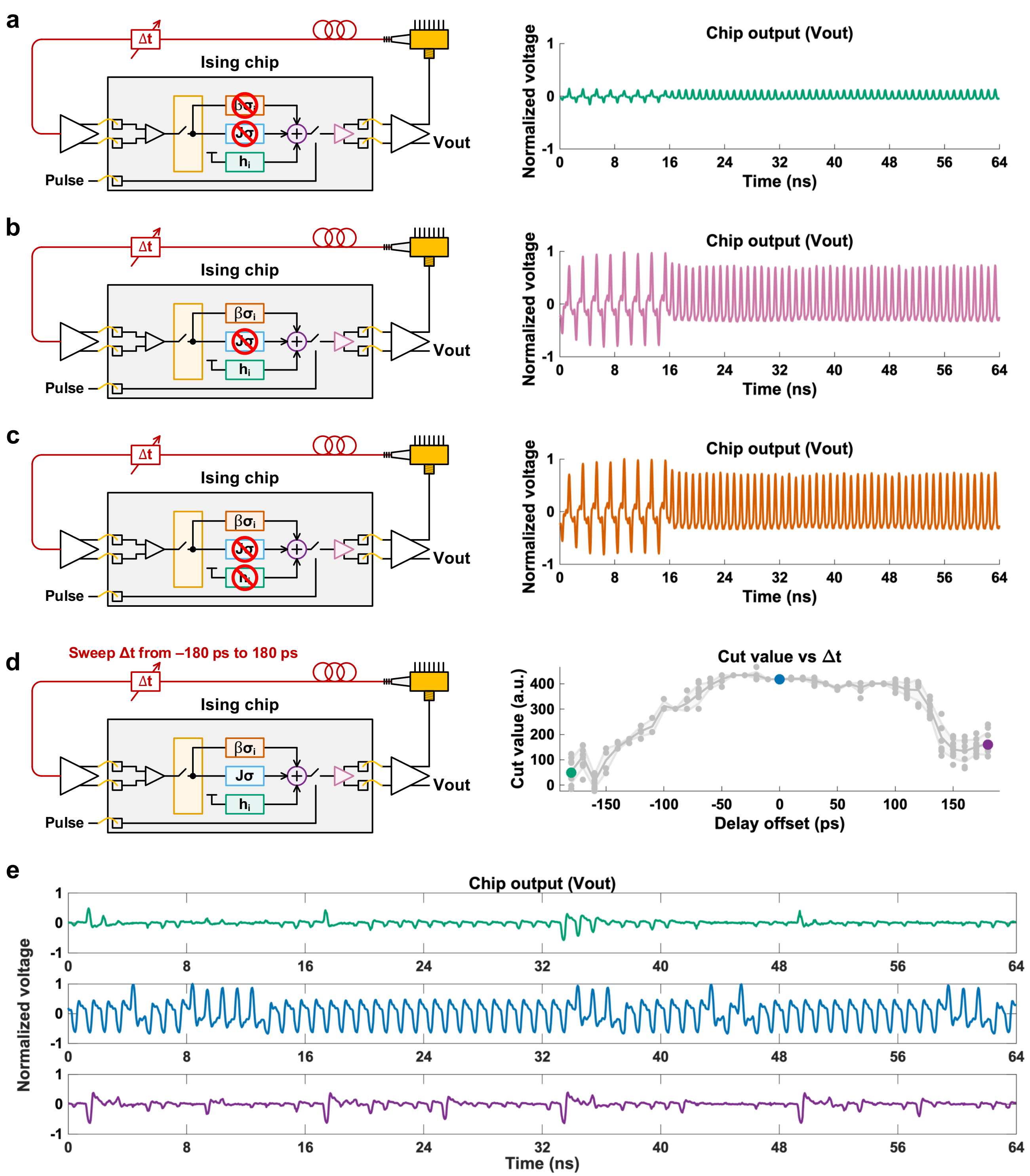

**Extended Data Fig. 5 | System timing alignment and sensitivity.** A calibration vector was used to verify the 64 ns round-trip time and ensure pulses going around the loop are sampled at the correct time by the chip. **a** First, with the feedforward and $\mathbf{J}\boldsymbol{\sigma}$ paths disabled, a test spin vector is launched into the optical delay line using the $\mathbf{h}$ vector. **b** Second, the feedforward path is enabled to amplify the pattern inside the optical delay line. **c** Third, the $\mathbf{h}$ vector is disabled, and the feedforward path sustains the pattern originally set by the $\mathbf{h}$ vector, indicating proper timing alignment. **d** To characterize the timing sensitivity of the system, the variable optical delay line is adjusted from -180 ps $\leq \Delta t \leq$ 180 ps in 10 ps increments. For each delay offset, a MaxCut problem is solved 10 times and the cut value is computed based on the solution generated by the chip. Mean (center line) and standard deviation (outer lines) of the produced cut values are included for reference. **e** For large delay offsets (green and purple), the chip is unable to correctly sample the pulses in the optical delay line and the chip output is significantly degraded compared to the case where the system is properly time aligned (blue).

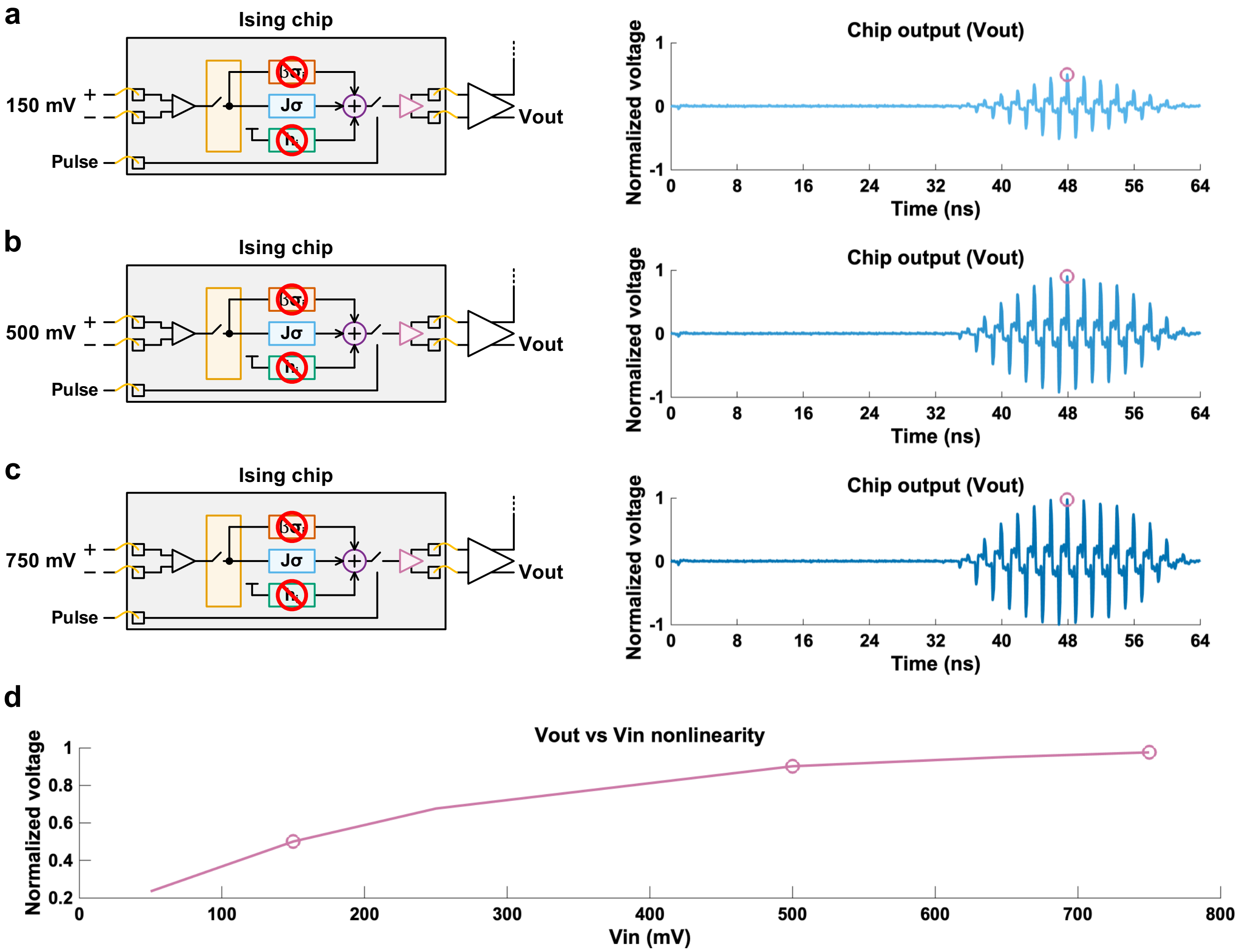

**Extended Data Fig. 6 | System nonlinearity characterization.** To characterize the nonlinearity of the system, dominated by the nonlinearity of the driver stage, the **J** matrix is programmed such that the last 32 spins exhibit an "alternating ramp" waveform (i.e. a sequence of alternating +1 and -1 spins with a triangular envelope). **a** The loop is opened, and a differential DC voltage with an amplitude of 150 mV is applied to the input of the chip. **b** As the differential input voltage increases to 500 mV, nonlinearity starts to appear. **c** With a differential input voltage of 750 mV, the alternating ramp saturates. **d** Plot of Vout vs Vin nonlinearity using the circled peak values from the alternating ramp waveforms.

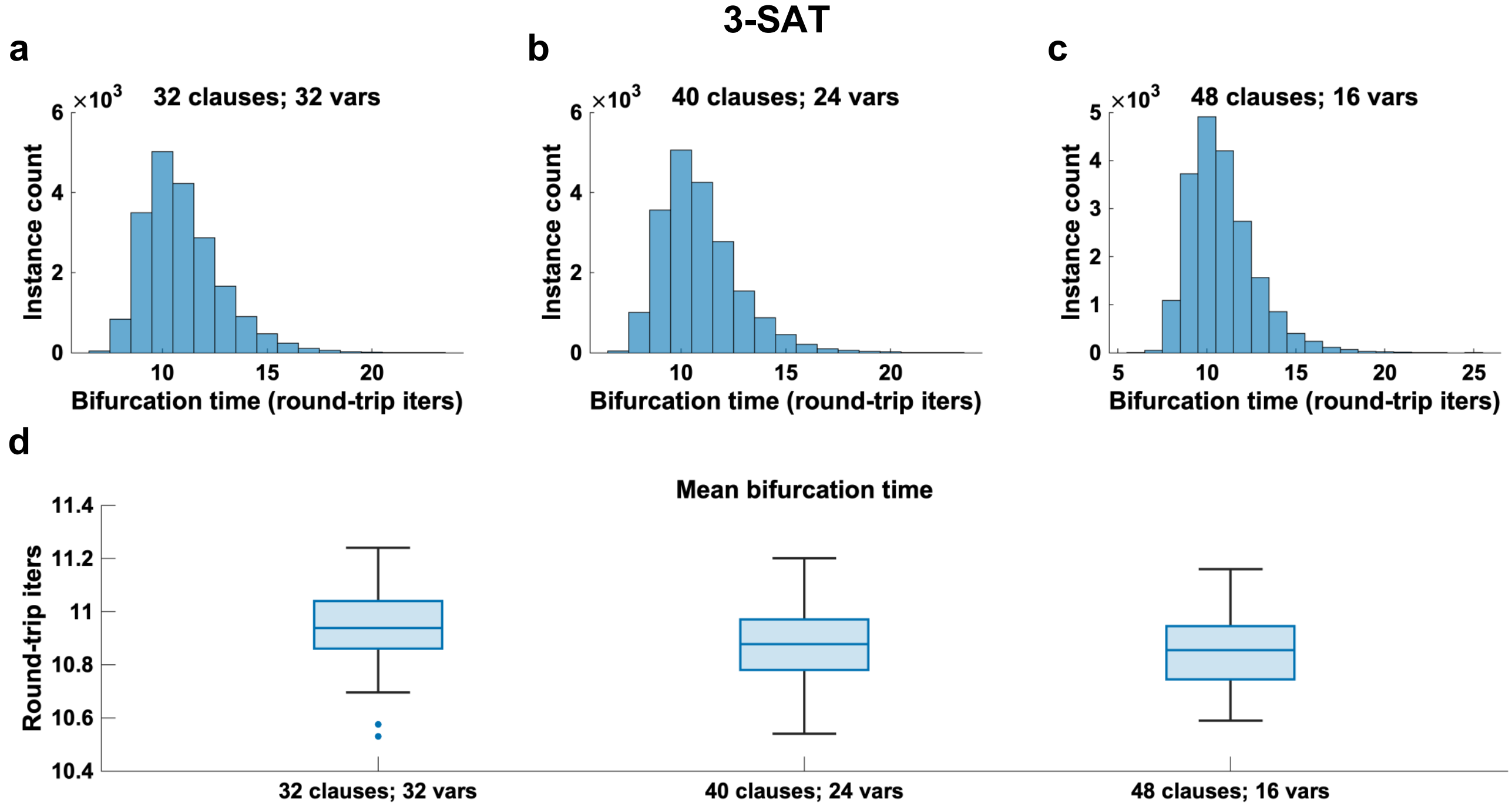

**Extended Data Fig. 7 | Ising problem time to solution.** Number of round-trip iterations for spins to bifurcate to 90% of their final value for 3-SAT problems with clause-to-variable ratios, $m/n$, of **a** 32/32, **b** 40/24, and **c** 48/16 with **d** associated box charts, indicating an average annealing time of approximately 700 ns.

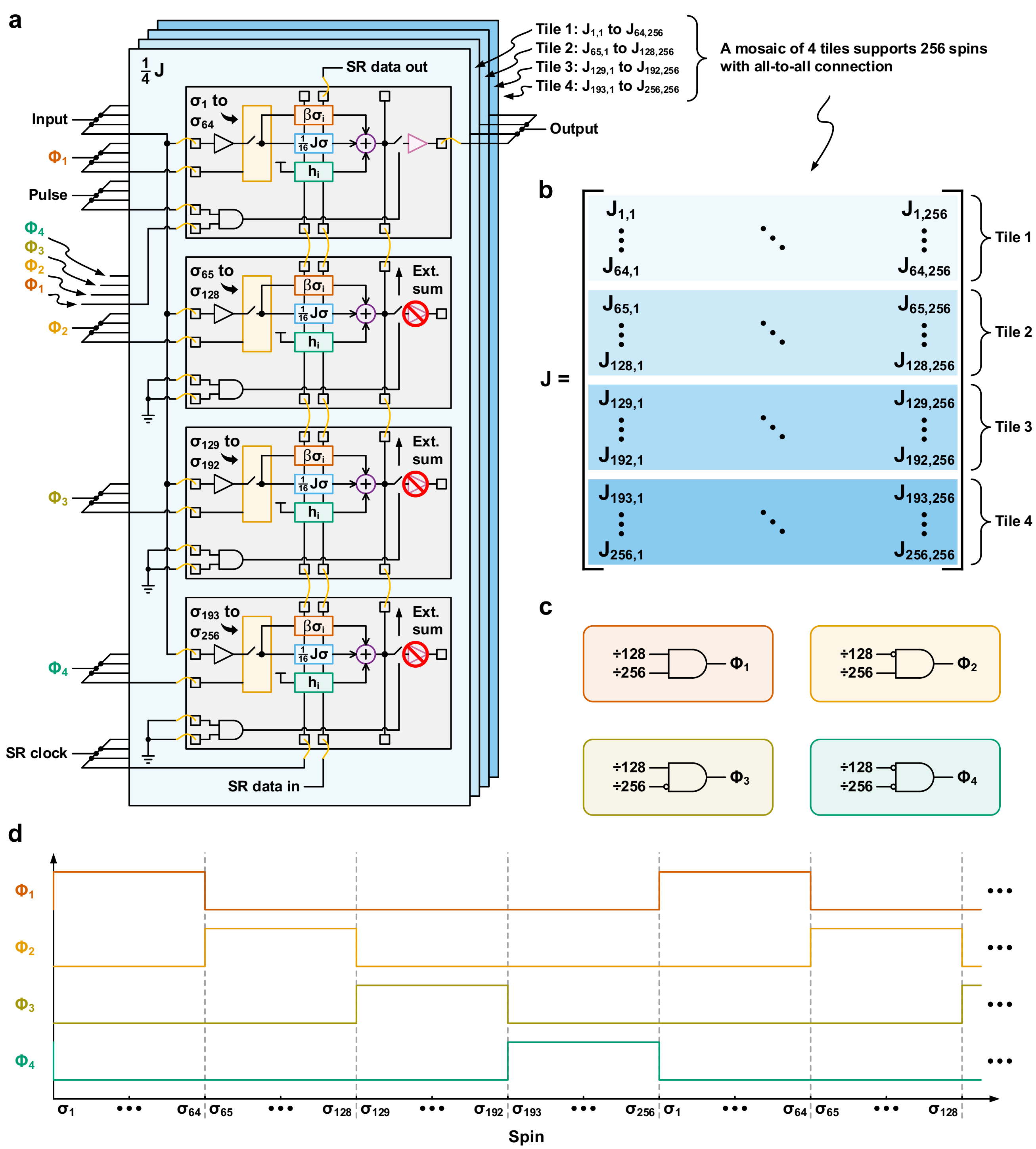

**Extended Data Fig. 8 | A path towards scalability.** The scalable nature of the design allows multiple chips to be connected to support larger spin counts. An example for 256 spins with all-to-all coupling is shown. **a** For 256 spins, a mosaic of 16 chips may be connected together in a set of 4 tiles, where each tile stores ¼ of the **J** matrix. **b** Breakdown of the **J** matrix for 256 spins. **c** Proposed circuitry that utilizes the divider outputs to generate the necessary clock waveforms as illustrated in **d**.

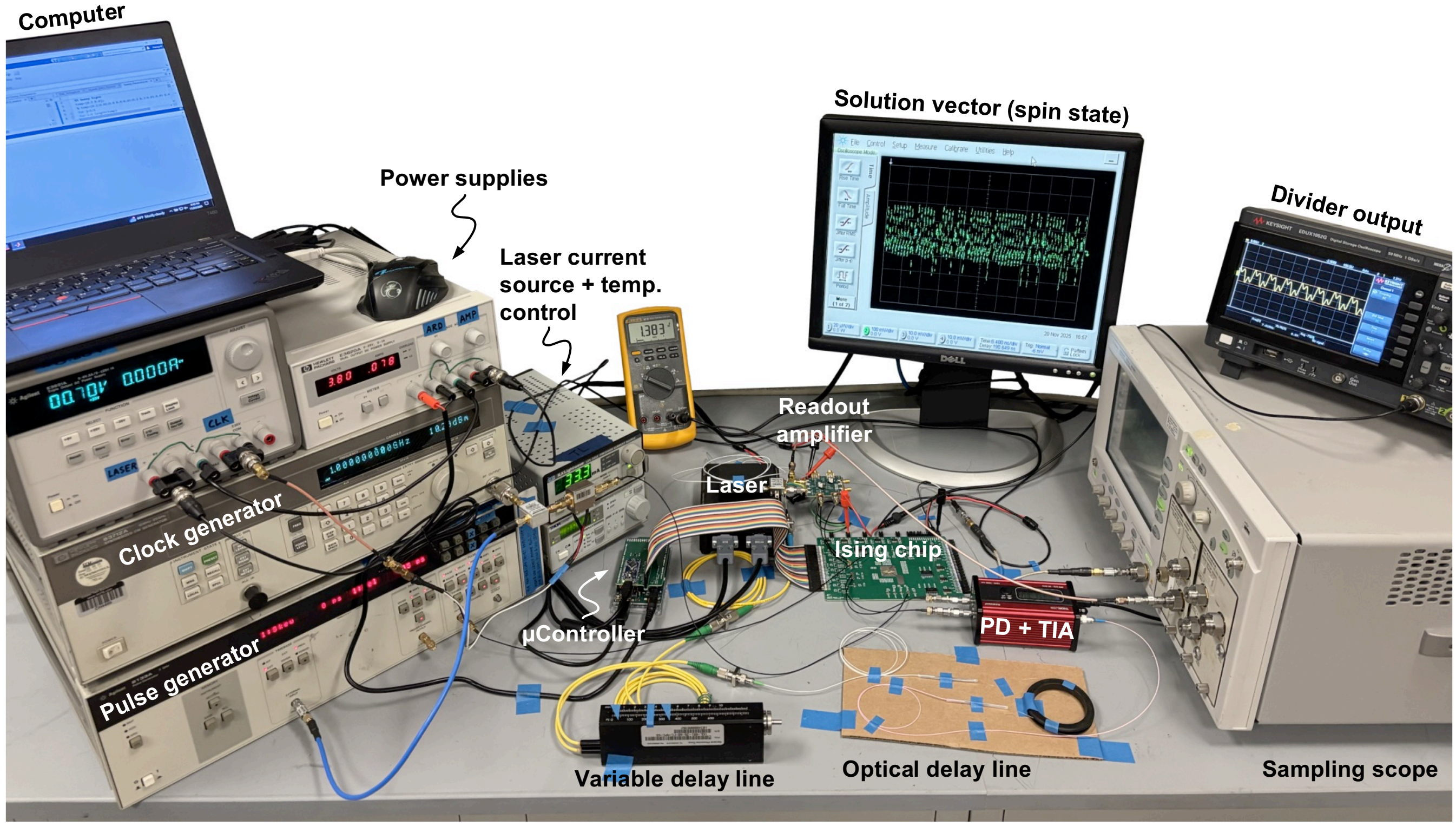

| Equipment | Model |
|---|---|
| DC power supply 1 | Agilent E3631A |
| DC power supply 2 | HP E3620A |
| Clock generator | HP 83712A |
| Pulse generator | HP 8133A |
| Laser | Mitsubishi FU-641SEA-1 |
| Laser current source | LDX-3412 |
| Laser temperature controller | TED 200 |
| Bias-Tee | Mini-Circuits ZX85-12G-S+ |
| Microcontroller | ATmega2560 (Arduino Mega 2560 R3) |
| Real-time oscilloscope | Keysight EDUX1052G |
| Sampling oscilloscope | Agilent DCA-J 86100C with HP 83484A & 83485A modules |
| Photodiode (PD) + transimpedance amplifier (TIA) | RXM25AF |
| Variable delay line | General Photonics VDL-001-35-60-FC/APC-SS |
| Readout amplifier | Analog Devices ADL5569-EVALZ |

**Extended Data Fig. 9 | Labeled measurement setup with a list of equipment and devices.**